\documentclass[useAMS,usenatbib]{mn2e}

\usepackage{graphicx}

\usepackage{booktabs}

\title [Relationship between H$_2$ and structure of PNe]{
On the relationship between the H$_2$ emission and the physical structure of
planetary nebulae}

\author[R.A. Marquez-Lugo et al.] 
{R.A.\ Marquez-Lugo$^{1}$\thanks{E-mail:
alejmar@astro.iam.udg.mx (RAML)}, 
G.\ Ramos-Larios$^{1}$, M.A.\ Guerrero$^{2}$, and R.\ V\'azquez$^{3}$ \\
$^{1}$Instituto de Astronom\'ia y Meteorolog\'ia, 
Av.\ Vallarta No.\ 2602, Col.\ Arcos Vallarta, 
C.P.\ 44130 Guadalajara, Jalisco, Mexico \\
$^{2}$Instituto de Astrof\'isica de Andaluc\'ia, IAA-CSIC, 
C/ Glorieta de la Astronom\'ia s/n, 
18008 Granada, Spain \\
$^{3}$Instituto de Astronom\'ia, Universidad Nacional Aut\'onoma de M\'exico, 
Apdo.\ Postal 877, 22800 Ensenada, B.C., Mexico 
}

\begin{document}

\date{2012 June 30}

\pagerange{\pageref{firstpage}--\pageref{lastpage}} \pubyear{2012}

\maketitle

\label{firstpage}

\begin{abstract}

Mid-IR observations of planetary nebulae (PNe) have revealed diffuse emission 
associated to their main nebular shells and outer envelopes or haloes.  
The interpretation of this emission is uncertain because the broad-band 
mid-IR images may include contributions of different components.  
In particular, the \emph{Spitzer} IRAC 8~$\mu$m images, that best  
reveal these nebular features, can include contributions not only 
of H$_2$ lines, but also those of ionic species, PAH features, and 
thermal dust continuum emission.  
To investigate the nature of the emission detected in mid-IR observations 
of a sample of 10 PNe, we have obtained narrow-band near-IR H$_2$ 
$\lambda$2.122 $\mu$m and optical [N~{\sc ii}] $\lambda$6584 \AA\ 
images. 
The comparison between these images confirm that a significant fraction 
of the emission detected in the IRAC 8~$\mu$m images can be attributed 
to molecular hydrogen, thus confirming the utility of these mid-IR images 
to investigate the molecular component of PNe.  
We have also detected H$_2$ emission from PNe whose physical structure 
cannot be described as bipolar, but rather as ellipsoidal or barrel-like. 
These detections suggest that, as more sensitive observations of PNe 
in the H$_2$ $\lambda$2.122 line are acquired, the detection of H$_2$ 
emission is not exclusive of bipolar PNe, although objects with this morphology
 are still the brightest H$_2$ emitters. 
Finally, we remark that the bright H$_2$ emission from the equatorial ring 
of a bipolar PN does not arise from a photo-dissociation region shielded 
from the UV stellar radiation by the ring itself, but from dense knots and 
clumps embedded within the ionized material of the ring.


\end{abstract}


\section{Introduction}

Molecular hydrogen (H$_2$) can be expected in the photodissociation 
regions (PDR) of planetary nebulae (PNe), where the expanding envelope 
sweeps up the wind from the progenitor asymptotic giant branch (AGB) 
star.  
Molecular hydrogen can also form in neutral clumps embedded 
within the ionization zone, where high extinction, high density, 
and molecular (and dust) shielding allow H$_2$ to survive from 
the stellar radiation \citep{2007ApJ...659.1291M}.
The H$_{2}$ molecules in PNe can be excited by the ultraviolet (UV) 
radiation field of their central stars (CSPNe) in the photodissociation 
front \citep{1987ApJ...322..412B} or by shocks \citep{1992ApJ...399..563B}. 
It has been suggested recently that H$_2$ can originally form 
in an excited state \citep{2011A&A...528A..74A}.

\begin{table*}
\begin{center}
\begin{minipage}{135mm}   
\caption{\label{tab:observations}Optical and infrared imaging.}
   \begin{tabular}{@{}lccccccccccc@{}}
   \hline
 & \multicolumn{2}{c}{Optical} & & &Near-IR& & & \multicolumn{2}{c}{Mid-IR}  \\
\cline{2-3}
\cline{5-7}
\cline{9-10}
            & [N~{\sc ii}] & $R$      & & H$_2$       & $K_s$       & $K_c$        & & $W2$ & IRAC4 \\
            & 6584 \AA     & 6600 \AA & & 2.122 $\mu$m & 2.17 $\mu$m & 2.27 $\mu$m & & 4.6 $\mu$m & 8 $\mu$m \\
  \hline
A\,66       & \dots        & DSS   & & NTT   & \dots & \dots & & \dots       & \emph{Spitzer} &~ \\
M\,1-79     & 1.50 m OAN          & \dots & & WHT   & \dots & WHT   & & \emph{WISE} & \dots          &~ \\
M\,2-48     & 1.50 m OAN   & \dots & & WHT   & \dots & WHT   & & \dots       & \emph{Spitzer} &~ \\
M\,2-51     & 1.50 m OAN   & \dots & & \dots & 2MASS & \dots & & \dots       & \emph{Spitzer} &~ \\
NGC\,650-51 & 0.84 m OAN   & \dots & & WHT   & \dots & WHT   & & \dots       & \emph{Spitzer} &~ \\
NGC\,6537   & 0.84 m OAN   & \dots & & WHT   & \dots & WHT   & & \dots       & \emph{Spitzer} &~ \\
NGC\,6563   & 1.50 m OAN   & \dots & & WHT   & \dots & WHT   & & \dots       & \emph{Spitzer} &~ \\
NGC\,6772   & 1.50 m OAN   & \dots & & WHT   & \dots & WHT   & & \dots       & \emph{Spitzer} &~ \\
NGC\,6778   & NOT          & \dots & & WHT   & \dots & WHT   & & \emph{WISE} & \dots          &~ \\
NGC\,7048   & 1.50 m OAN   & \dots & & TNG   & \dots & TNG   & & \dots       & \emph{Spitzer} &~ \\
 \hline
 \end{tabular}
\textbf{Notes:}
2MASS (Two Micron All Sky Survey), 
DSS (Digitized Sky Survey), 
NOT (Nordic Optical Telescope), 
NTT (New Technology Telescope), 
OAN (Observatorio Astron\'omico Nacional),  
\emph{Spitzer} (\emph{Spitzer Space Telescope}), 
TNG (Telescopio Nazionale Galileo),  
WHT (William Herschel Telescope), 
\emph{WISE} (\emph{Wide-field Infrared Survey Explorer}). 
\end{minipage}
\end{center}
\end{table*}

It has been traditionally assumed that H$_2$ emission arises from 
regions where material is predominantly molecular. 
Therefore, it is not surprising that H$_2$ emission is mainly 
detected at the equatorial regions of bipolar PNe\footnote{
Hereafter we will adopt the definition of bipolar PNe given by Corradi \&
Schwarz (1995) as those whose H$\alpha$ images display ``{\it an equatorial waist from which
two faint, extended bipolar lobes depart}''.
When the morphology of the PN were insufficient to determine its
physical structure (a ring-like PN can be interpreted as a pole-on
bipolar source), we will rely only on kinematical information of
the source.
}, because their thick equatorial disks would shield the UV radiation of the PN central star (CSPN), allowing 
the survival of H$_2$ molecules \citep{1996ApJ...462..777K,2000ApJS..127..125G}. 
Furthermore, the confinement of bipolar PNe to low Galactic latitudes 
have made them suspected to descend from the population of most massive 
progenitor stars of PNe \citep{1995A&A...293..871C}.  
As these stars would eject thicker and more massive envelopes, this 
is considered an additional argument linking bipolar morphology with 
the significant presence of molecular material 
\citep{1983IAUS..103..233P,1996AAS...189.9708H}.
The strong correlation between H$_2$ emission and bipolar morphology
 has originated the so-called 
\emph{Gatley's rule} \citep{1994ApJ...421..600K}: ``the detection of 
the 2.122 $\mu$m S(1) line of H$_2$ is sufficient to determine the 
bipolar nature of a PN."

The increase in sensitivity of near- and mid-IR observations of H$_2$ lines 
\citep{2006IAUS..234..173H}, and the access to other wavelength ranges (e.g., 
far-UV by \emph{FUSE}, the Far-UV Spectroscopic Explorer) have revealed the 
presence of molecular hydrogen in PNe with a variety of morphologies \citep{2006ASPC..348..328D}. 
The wavelength range between 1 and 10 $\mu$m is especially 
relevant because it includes a large number of 
transitions of molecular hydrogen \citep{1977BAAS....9Q.591T}.  
The intensity line ratios of some of these lines (the 2--1~S(1)~$\lambda$2.2477/1--0~S(1)~$\lambda$2.1218
and 1--0~S(0)~$\lambda$2.2235/1--0 S(1)~$\lambda$2.1218 particularly) can be used to infer the molecular 
excitation mechanism (shocks or UV radiation field) and the physical 
conditions of H$_2$ \citep[see][]{2006AJ....131.1515L}.

The four \emph{Spitzer} Infrared Array Camera (IRAC) bands include 
H$_2$ lines such as the 1--0 O(5) $\lambda$3.235 $\mu$m and 0--0 S(13) 
$\lambda$3.8464 $\mu$m lines in the IRAC 3.6 $\mu$m band, the 0--0 S(1) 
$\lambda$4.181 $\mu$m and 0--0 S(9) $\lambda$4.6947 $\mu$m lines in the 
IRAC 4.5 $\mu$m band, the 0--0 S(7) $\lambda$5.5115 $\mu$m and 0--0 S(6) 
$\lambda$6.1088 $\mu$m lines in the IRAC 5.8 $\mu$m band, and the 0--0 
S(5) $\lambda$6.9091 $\mu$m and 0--0 S(4) $\lambda$8.0258 $\mu$m lines 
in the IRAC 8 $\mu$m band. 
Observations of PNe have shown the relevance of nebular 
emission in the IRAC 8 $\mu$m band as it contains few emission 
from stars and more intense diffuse emission than the IRAC 3.6 
and 4.5 $\mu$m bands \citep[e.g.,][]{2012A&A...537A...1A}, 
while having a better sensitivity than the IRAC 5.8 $\mu$m band.
Furthermore, the IRAC 3.6 and 4.5 $\mu$m bands are dominated by bremsstrahlung 
emission and by the ionic transition lines of Br$\alpha$ $\lambda$4.052 
$\mu$m, [Mg~{\sc iv}] $\lambda$4.49 $\mu$m, and [Ar~{\sc vi}] $\lambda$4.53 
$\mu$m \citep{2010MNRAS.405.2179P}. 
IRAC 8 $\mu$m images of PNe frequently reveal 
emission from extended haloes and molecular knots 
\citep{2009MNRAS.400..575R,2011arXiv1109.4439C,2011MNRAS.410.2257P}.  
The comparison between IRAC 8 $\mu$m and H$_2$ images of PNe (e.g. NGC\,6720 and NGC\,7293) reveals an excellent match between 
the appearance in the two bands that indicates a common origin 
\citep{2006IAUS..234..173H,2006ApJ...652..426H}, although molecular hydrogen is suspected, the contribution from the [Ar~{\sc iii}] 
$\lambda$8.991 $\mu$m line cannot be neglected \citep{2004ApJS..154..296H}.

Accordingly we have searched the \emph{Spitzer} archive for IRAC 
8 $\mu$m images of PNe showing extended emission that could be 
attributed to molecular material. The sample, composed of 8 morphologically
diverse PNe, has subsequently been imaged through narrow-band optical
and near-IR filters to ascertain the presence and spatial distribution
of ionized material and H$_2$.  Two PNe, namely M\,1-79 and NGC\,6778,
were added to this sample as they exhibit intriguing narrow-band optical
and near-IR morphologies, despite there are no \emph{Spitzer} images for
them.  The comparison between optical, near- and mid-IR images has allowed us to 
verify the presence of molecular material in regions of H$_2$ emission, but also from regions where H$_2$ 
is shielded from the stellar UV radiation field, so that the H$_2$ molecules 
are neither disrupted nor excited. 
In this paper we present evidence that the bipolar PNe--H$_2$ relationship 
is not as close as claimed by Gatley's rule, as H$_2$ emission is found in 
PNe with morphological types other than bipolar.

The observations and archival data are presented in Section \ref{sec_obs}. 
The results for each individual PN are described in Section \ref{sec_res}.
The discussion and final summary are presented in Section \ref{sec_dis}.

\section[]{Observations and Archival Data}\label{sec_obs}

\subsection{Optical images}

\begin{figure*}
\begin{minipage}{175mm}   
\begin{center}
\includegraphics[width=80mm]{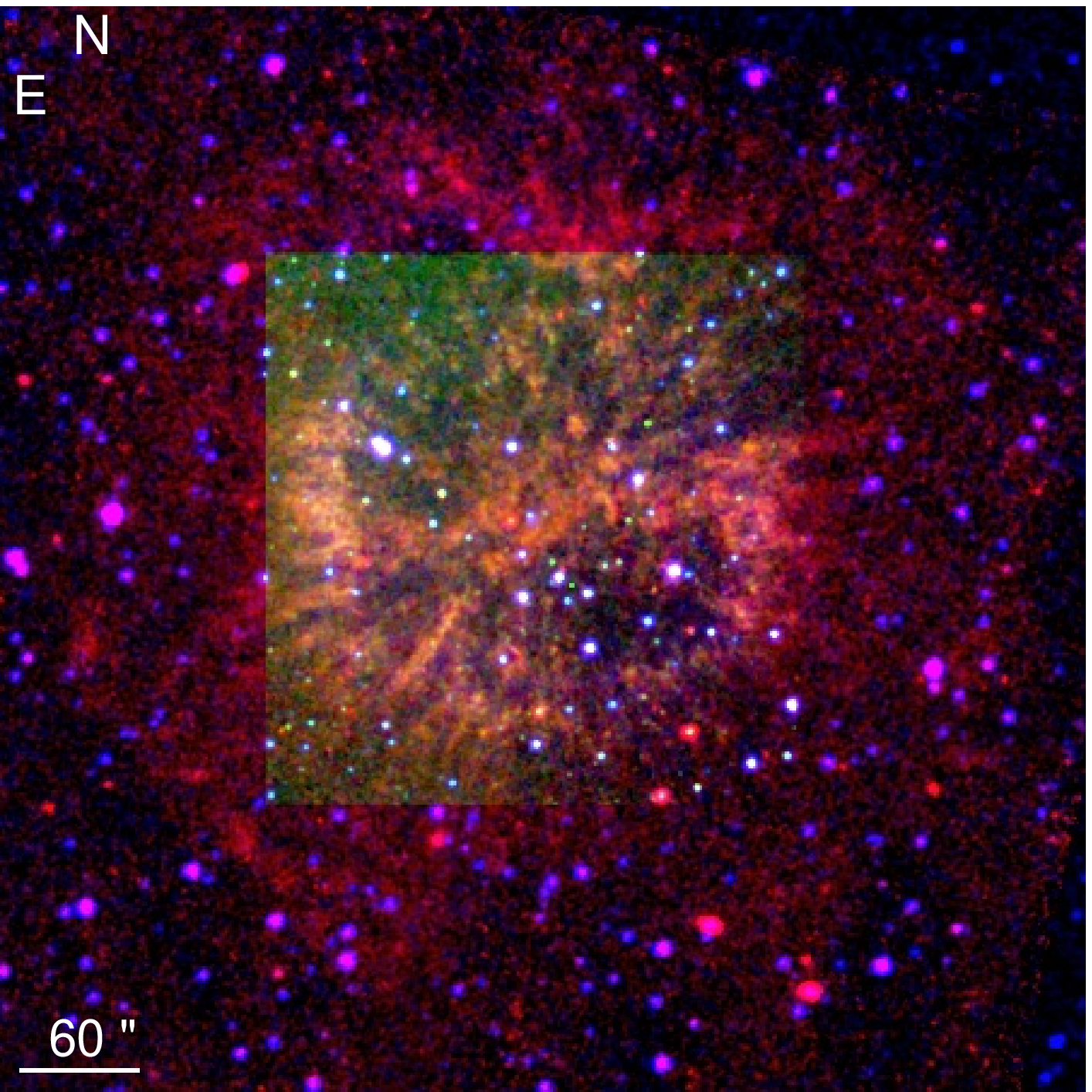}
\hspace*{\columnsep}
\includegraphics[width=80mm]{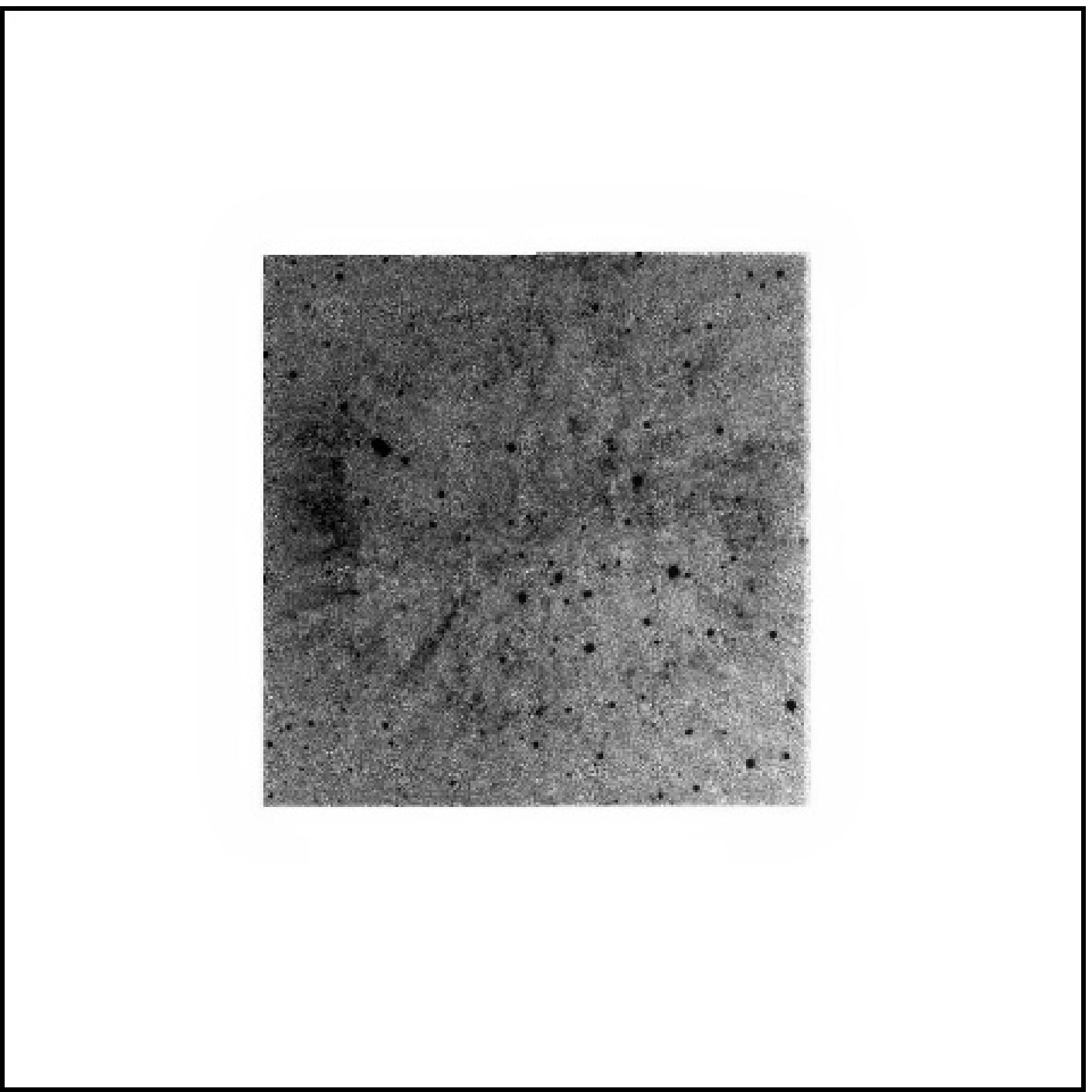}
\caption{
IRAC 8~$\mu$m (red), NTT H$_2$ $\lambda$2.122 (green), and DSS $R$ (blue) 
colour-composite RGB picture ({\it left}), and NTT H$_2$ $\lambda$2.122 
image ({\it right}) of A\,66. Note the fringing that affects the Southwest corner of the H$_2$ image.
}
\label{A66}
\end{center}
\end{minipage}
\end{figure*}

Most of the optical images presented in this paper 
(Table~\ref{tab:observations}) have been obtained 
at the Observatorio Astron\'omico Nacional (OAN, 
Mexico), using either the 1.5m Harold Johnson or the 0.84m telescopes. 
At the 1.5m telescope, a [N~{\sc ii}] filter ($\lambda_c$=6584 \AA, 
$\Delta\lambda$=11 \AA) was used, whereas at the 0.84m telescope 
either a [N~{\sc ii}]+H$\alpha$ filter ($\lambda_c$6564 \AA, 
$\Delta\lambda$=72 \AA) or a [N~{\sc ii}] filter ($\lambda_c$6583 \AA, 
$\Delta\lambda$=10 \AA) were used. 
The narrow-band [N~{\sc ii}] images of M\,1-79, M\,2-48, M\,2-51, and NGC\,7048 were obtained 
using the RUCA filter wheel \citep{2000RMxAA..36..141Z} at the OAN 
1.5m telescope. 
The detector was the SITe1 1024$\times$1024 CCD with pixel scale 
0\farcs252 pixel$^{-1}$, binning 2$\times$2 and FoV 4\farcm2$\times$4\farcm2.
The narrow-band [N~{\sc ii}] images of NGC\,6563 and NGC\,6772 were obtained
using the RUCA filter wheel at the OAN 1.5m telescope with the detector Marconi e2v 2048$\times$2048 CCD with pixel scale 
0\farcs14 pixel$^{-1}$, binning 1$\times$1 and 2$\times$2 respectively and FoV 4\farcm7$\times$4\farcm7.
The narrow-band [N~{\sc ii}] image of NGC\,6537 was obtained using the 
MEXMAN filter wheel at the 0.84m 
OAN telescope with the SITe4 1024$\times$1024 CCD that provides 
a pixel scale of 0\farcs39 pixel$^{-1}$, binning 1$\times$1 and a FoV of 6\farcm7$\times$6\farcm7. 
The [N~{\sc ii}]+H$\alpha$ image of NGC\,650-51 was obtained 
using SOPHIA (Sistema \'Optico Para Hacer Im\'agenes de campo Amplio), 
an optical system for the acquisition of wide-field images at the 0.84m 
OAN telescope. 
This time, the detector was a 2048$\times$4608 CCD e2v, named ESOPO, 
providing a pixel scale of 1\farcs07 pixel$^{-1}$, binning 1$\times$1 and a FoV of $\simeq$30\farcm0$\times$30\farcm0. 

For NGC\,6778, we have used the [N~{\sc ii}] images 
published by \citet{2010PASA...27..180M}.  
The images were obtained through the narrow-band [N~{\sc ii}] 
filter ($\lambda_c$=6584 \AA, $\Delta\lambda$=9 \AA) using 
ALFOSC (Andalucia Faint Object Spectrograph and Camera) in its 
imaging mode at the 2.56m Nordic Optical Telescope (NOT) of the 
Observatorio del Roque de los Muchachos (ORM), La Palma, Spain. 
The e2v 2048$\times$2048 CCD detector used for these observations 
provides a pixel scale of 0\farcs184 pixel$^{-1}$ and a FoV of 
6\farcm3$\times$6\farcm3.

Finally, for A\,66 we used an $R$ band POSS-II image downloaded from 
the ESO Digitized Sky Survey\footnote{
The DSS is an all-sky photographic survey 
conducted with the Palomar and UK Schmidt telescopes. 
The Catalogs and Surveys Branch (CASB) is digitizing the 
6\fdg5$\times$6\fdg5 photographic plates using a modified 
PDS microdensitometer to support \emph{HST} observing 
programs and to provide a service to the astronomical 
community. } 
(DSS) with pixel scale $\simeq$1\farcs0 pixel$^{-1}$.

\subsection{Near-IR images}

The narrow-band H$_2$ $\lambda$2.122~$\mu$m and $K_c$ continuum near-IR 
images were mainly obtained at the 4.2m William Herschel Telescope (WHT) 
of the ORM using LIRIS \citep[Long-Slit Intermediate Resolution Infrared 
Spectrograph,][]{2003INGN....7...15A}. 
The detector was a 1024$\times$1024 HAWAII array with plate scale 
0\farcs25 pixel$^{-1}$ and FoV of 4\farcm3$\times$4\farcm3. 
The narrow-band H$_2$ $\lambda$2.122~$\mu$m and $K_S$ continuum 
$\lambda$2.27~$\mu$m images of NGC\,7048 were obtained 
using NICS 
\citep[Near Infrared Camera and Spectrometer,][]{1995A&AS..114..179O} 
at the 3.5m Telescopio Nazionale Galileo (TNG) of the ORM. 
The Rockwell Hawaii 1024$\times$1024 array used for these 
observations has a projected scale of 0\farcs25 pixel$^{-1}$ 
and a FoV of 4\farcm3$\times$4\farcm3. 
The narrow-band H$_2$ $\lambda$2.122~$\mu$m image of A\,66 was obtained 
using the SOFI camera \citep[Son OF Isaac,][]{1998Msngr..91....9M} at 
the 3.5m New Technology Telescope (NTT) of La Silla Observatory, Chile. 
The detector was a 1024$\times$1024 HgCdTe HAWAII array providing in its 
large field mode (LF) a pixel scale of 0\farcs288~pixel$^{-1}$ and a FoV 
of 4\farcm9$\times$4\farcm9.  
Finally, Two Micron All Sky Survey (2MASS) \emph{JHK$_s$} images with 
pixel scale 1\farcs0~pixel$^{-1}$ are presented for M\,2-51.


\subsection{Mid-IR images}

\begin{figure*}
\begin{minipage}{175mm}   
\begin{center}
\includegraphics[width=80mm]{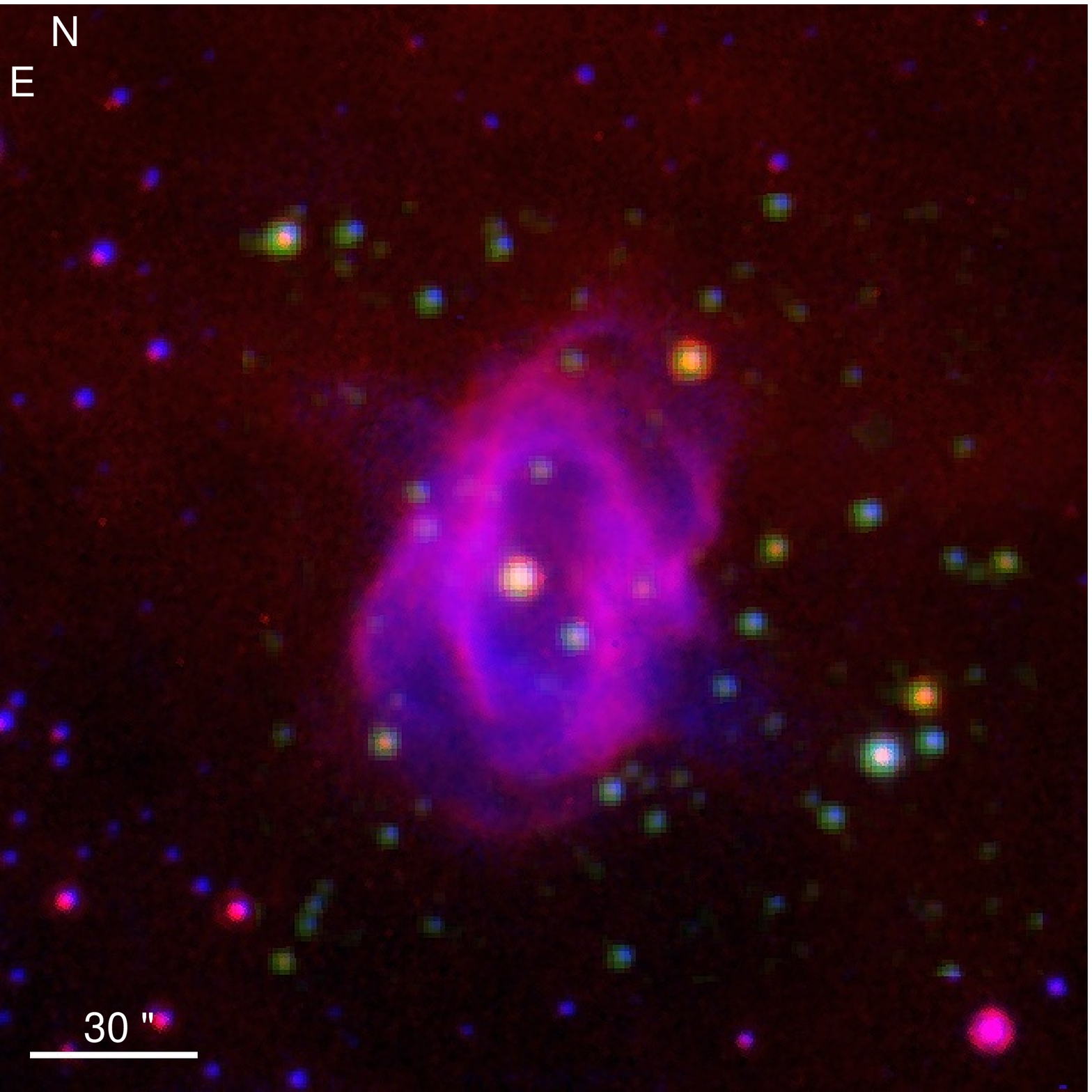}
\hspace*{\columnsep}
\includegraphics[width=80mm]{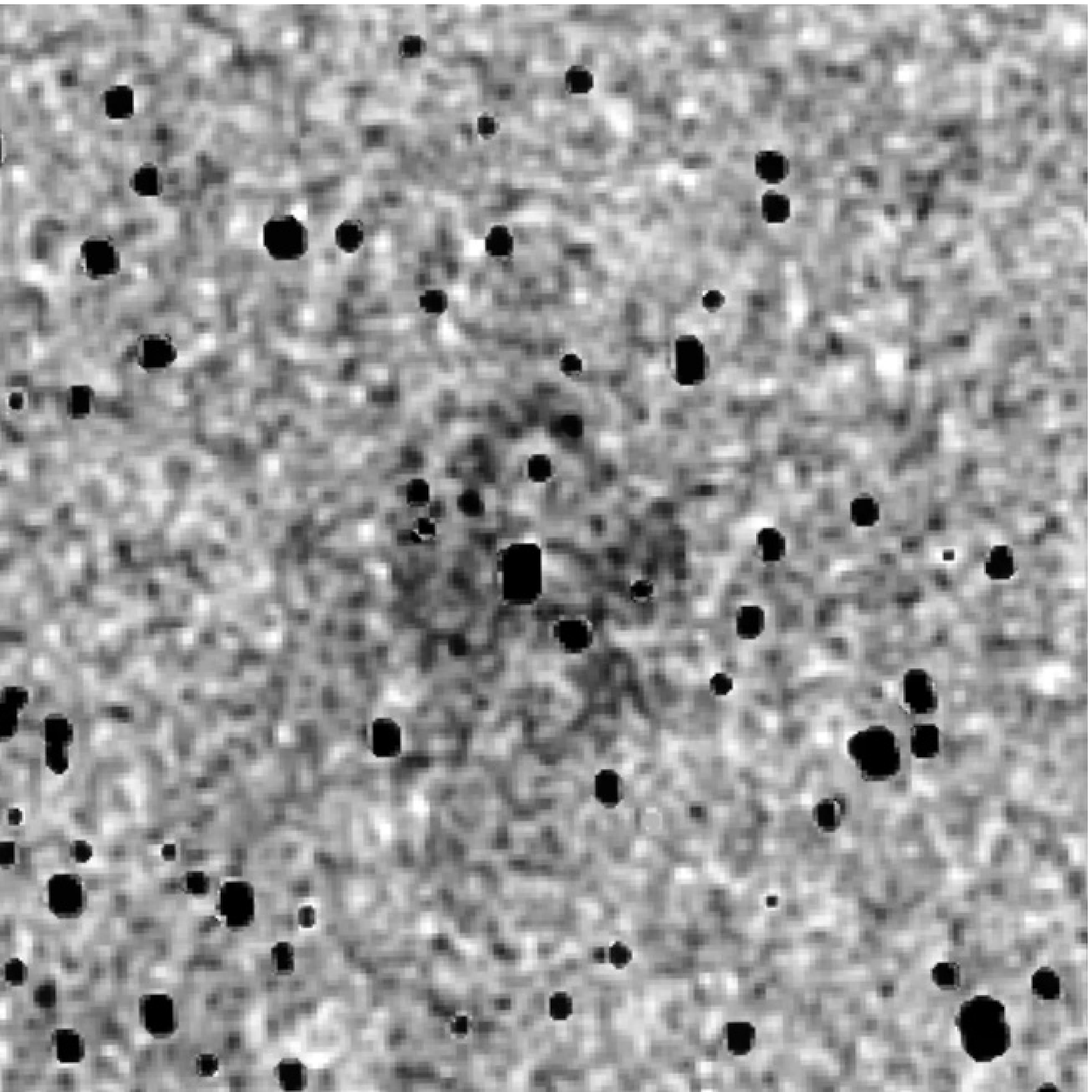}
\caption{
IRAC 8~$\mu$m (red), 2MASS K$_s$ $\lambda$2.17 (green), and OAN [N~{\sc ii}] 
(blue) colour-composite RGB picture ({\it left}), and 2MASS $K_s$ excess image ({\it right}) of M\,2-51. 
}
\label{M251}
\end{center}
\end{minipage}
\end{figure*}

The mid-IR images used in this paper have been downloaded mostly from the 
\emph{Spitzer} archives. 
The \emph{Spitzer} IRAC images of M\,2-48 and NGC\,6537 belong to the 
Galactic Legacy Infrared Mid-Plane Survey Extraordinaire (GLIMPSE) program 
which used IRAC to map the Galactic plane in the range $|l| \leq 60\degr, 
|b| \leq 1^{\circ}$ \citep{2004ApJS..154...10F}. 
The GLIMPSE images in the IRAC 8~$\mu$m band used in this paper have 
a spatial resolution $\simeq$2\arcsec. 
Similarly, we have used the \emph{Spitzer} IRAC 8 $\mu$m images of A\,66 
and NGC\,7048 (Program ID 30285, \emph{Spitzer} Observations of Planetary 
Nebulae 2, PI: Giovanni Fazio), NGC\,650-51, NGC\,6563, and NGC\,6772 
(Program ID 68, Studying Stellar Ejecta on the Large Scale using SIRTF-IRAC, 
PI: Giovanni Fazio), and M\,2-51 (Program ID 50398, \emph{Spitzer} Mapping of 
the Outer Galaxy, SMOG, PI Sean Carey).
The spatial resolution of these images varies between $\simeq$1\farcs7 and $\simeq$2\farcs0.  

No \emph{Spitzer} images are available for M\,1-79 and NGC\,6778.  
For these nebulae, we have used \emph{WISE} \citep[\emph{Wide-field Infrared 
Survey Explorer},][]{2010AJ....140.1868W} images retrieved from the NASA/IPAC 
Infrared Science Archive (IRSA). 
\emph{WISE} is a NASA Explorer mission to surveys the entire sky at 
3.4, 4.6, 12, and 22~$\mu$m, the so-called W1 through W4 bands, with 
5$\sigma$ point source sensitivities better than 0.08, 0.11, 1, and 6 
mJy, respectively. 
The 40cm telescope uses HgCdTe and Si:As detectors arrays with a plate 
scale of 2\farcs75 pixel$^{-1}$. 
The W2 4.6~$\mu$m images were downloaded from the WISE All-Sky Data 
Release.  
The images have angular resolution $\simeq$6\farcs4 and astrometric 
accuracy for bright sources better than 0\farcs15.

\subsection{Spitzer Spectroscopy in the MIR}

Spectroscopic observation used in this paper were acquired using the Short-Low
(SL) module 1 (SL1) and module 2 (SL2) at short
(5.1-8.5~$\mu$m) and long (7.4-14.2~$\mu$m) wavelength  respectively of the
Spitzer Infrared Spectrograph \citep[IRS;][]{2004SPIE.5487...62H}. The spectra of
M\,2-51 and NGC\,6537 were obtained through Spitzer Program 45 (Deuterium
Enrichment in PAHs; P.I. Thomas Roellig) on 1/6/2004 and Spitzer Program 50179
(Planetary Nebulae As A Laboratory For Molecular Hydrogen in the Early Universe; P.I.
Kris Sellgren) on 4/11/2008  respectively. 

\section{Results}\label{sec_res}

Based on the inspection of the optical images and previous 
spatio-kinematical studies, when available, we have divided 
our sample into two broad morphological groups: 
elliptical PNe and bipolar PNe.
The first group includes A\,66, M\,2-51, NGC\,6563, NGC\,6772, and NGC\,7048, 
whereas the group of bipolar PNe is composed of M\,1-79, M\,2-48, NGC\,650-51, 
NGC\,6537, and NGC\,6778.

\subsection{Elliptical PNe}\label{round}

\subsubsection{A\,66 --- PN\,G019.8$-$23.7}\label{A66s}

A\,66 was included in the list of old, evolved PNe compiled by 
\citet{1955PASP...67..258A,1966ApJ...144..259A}. 
The DSS optical image (blue in Figure~\ref{A66}-{\it left}) shows 
a roughly spherical, low surface brightness shell of radius $\simeq$118\arcsec.
The best quality narrow-band optical images of A\,66 were presented by 
\citet{1998A&AS..133..361H} who described it as a roundish, old PN with 
a blowout structure towards the NE.  
They also reported the presence of radial structures or filaments 
escaping outwards, and a band of emission in H$\alpha$ and [N~{\sc ii}] 
crossing the central regions and dividing the nebula into two cavities.  

\begin{figure*}
\begin{minipage}{175mm}   
\begin{center}
\includegraphics[width=80mm]{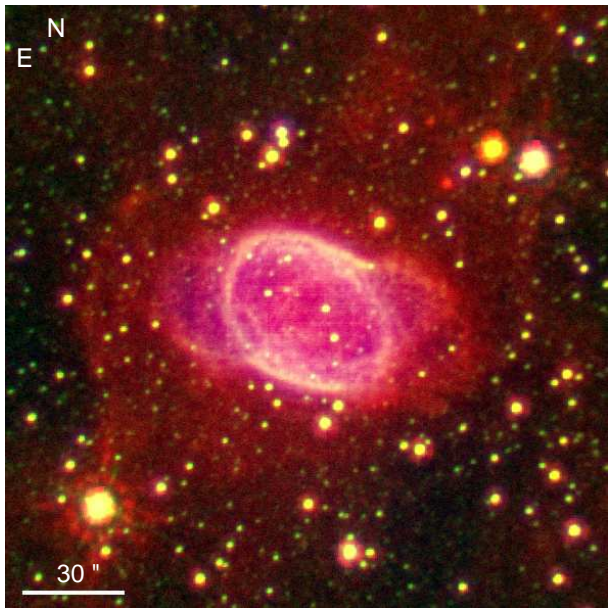}
\hspace*{\columnsep}
\includegraphics[width=80mm]{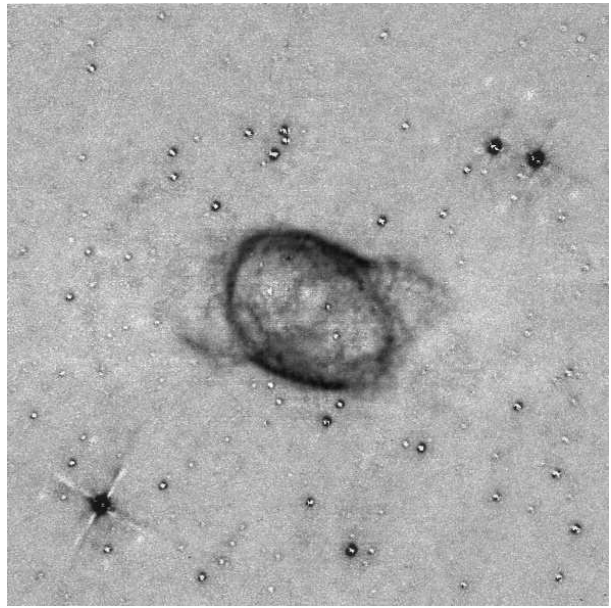}
\caption{ 
IRAC 8~$\mu$m (red), WHT H$_2$ $\lambda$2.122 (green), and OAN [N~{\sc ii}] (blue) 
colour-composite RGB picture ({\it left}), and WHT continuum-subtracted H$_2$ 
$\lambda$2.122 image ({\it right}) of NGC\,6563. 
}
\label{NGC6563}
\end{center}
\end{minipage}
\end{figure*}

The morphology of A\,66 in the H$_2$ and IRAC 8~$\mu$m images 
highlights the radial structures hinted in optical images.  
The H$_2$ image (Figure~\ref{A66}-{\it right}) shows a series of cometary 
knots that are mostly distributed along a central band and a fragmented 
ring of radius 133\arcsec\ broken towards the northeast.  
There is a clear correlation between these morphological features and 
those described in optical images: the central band is coincident with 
that observed in the optical, the ring-like structure encompasses the 
optical emission, and the lack of H$_2$ emission towards the northeast corresponds 
with the H$\alpha$ and [N~{\sc ii}] blowout.

The IRAC 8~$\mu$m image (red in Figure~\ref{A66}-{\it left}) shows very 
similar morphology in the central regions to that of the H$_2$ image, but 
its larger FoV reveals emission extending farther out.  
Indeed, the H$_2$ cometary knots that overrun the NTT H$_2$ image 
stretch out in the IRAC 8~$\mu$m emission up to radial distances 
$\simeq$240\arcsec.  
The central optical nebula is surrounded by a halo 
of emission in the IRAC 8~$\mu$m image.  
The similarities between H$_2$ and IRAC 8~$\mu$m emission in the central 
regions and the identification of some of the outermost features in the 
IRAC 8 $\mu$m image with radial knots in the H$_2$ image strongly suggests 
that the outermost emission detected in the IRAC 8~$\mu$m image is produced 
by H$_2$ molecules.

\subsubsection{M\,2-51 --- PN\,G103.2$+$00.6 }\label{M251s}

Optical [N~{\sc ii}] images of M\,2-51
\citep{1986ApJ...302..727J,1987AJ.....94..671B}
have revealed an elliptical morphology with a size
$\simeq$36\arcsec$\times$56\arcsec, and major
axis almost along the north-south direction.
Our [N~{\sc ii}] image detects this elliptical shell, as well 
as an external elliptically shaped outer shell of size 
$\simeq$60\arcsec$\times$86\arcsec whose major axis is tilted by 
$\simeq$30\degr\ with respect to that of the inner shell. 
Low surface brightness diffuse emission is also detected along
the minor axis of the outer elliptical shell up to a radial
distance of 40\arcsec.  \\

No narrow-band observations of the H$_2$ $\lambda$2.122~$\mu$m line are 
available in the literature for M\,2-51.  
We have compared the 2MASS $K_s$ image with those in the $J$ and
$H$ bands to search for a photometric excess that could be used 
as a proxy for detection of H$_2$ emission 
\citep[see][for details on this technique]{2006RMxAA..42..131R}.
The 2MASS image (Figure~\ref{M251}-{\it right}), similar in quality to the 
$JHK$ images presented by \citet{1999PASJ...51..673S}, reveals hints of 
emission excess from a filamentary structure consistent in size and location 
with the outer elliptical shell, but the low signal of this emission does not 
allow us to make a firm statement.  \\

The image in the IRAC 8~$\mu$m band (red in Figure~\ref{M251}-{\it left})
generally follows the double shell morphology and diffuse emission hinted 
in the [N~{\sc ii}] image.
The outer elliptical shell has a size $\simeq$80\arcsec$\times$116\arcsec.
The ionic or molecular nature of this emission is uncertain.  \\

\subsubsection{NGC\,6563 --- PN\,G358.5$-$07.3 }\label{NGC6563s}

NGC\,6563 was included in the catalogue of narrow-band images of PNe
of \citet{1992A&AS...96...23S}.  
Its H$\alpha$+[N~{\sc ii}] image displays a main body with elliptical shape 
and two extensions or {\it ansae}.  
Our [N~{\sc ii}] image (blue in Figure~\ref{NGC6563}-{\it left}) shows similar
elliptical morphology oriented along the northeast-southwest direction (major 
axis along P.A.$\simeq$50\degr) with a size of 38\arcsec$\times$52\arcsec.  
The two {\it ansae} protrude from the bright inner shell almost along 
the east-west direction up to radial distances $\simeq$40\arcsec.  
Whereas the optical morphology of NGC\,6563 may be interpreted as
a wide equatorial belt and narrow bipolar lobes seen almost pole-on,
the kinematics do not to confirm this interpretation (V\'azquez et
al., in prep.), but rather confirm an ellipsoidal structure with
short, low velocity extensions consistent with {\it ansae}
\citep{1993A&A...276..463S}.

\begin{figure*}
\begin{minipage}{175mm}   
\begin{center}
\includegraphics[width=80mm]{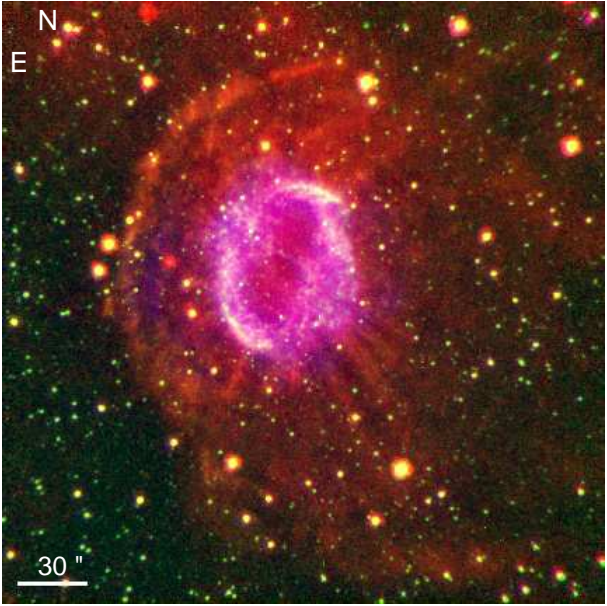}
\hspace*{\columnsep}
\includegraphics[width=80mm]{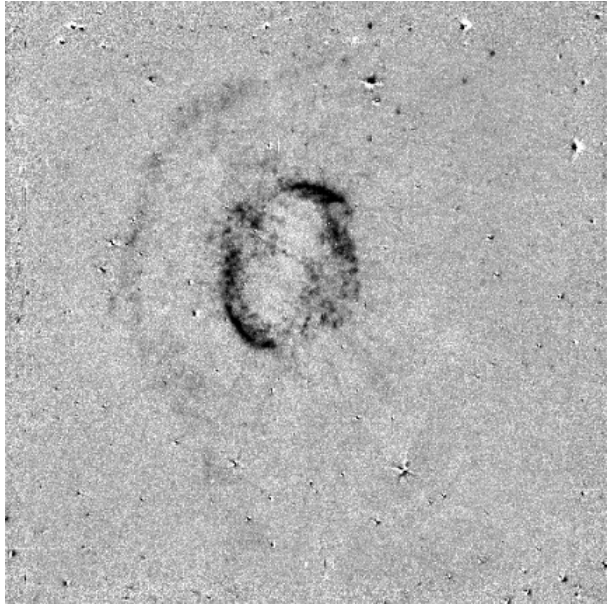}
\caption{ 
IRAC 8~$\mu$m (red), WHT H$_2$ $\lambda$2.122 (green), and OAN [N~{\sc ii}] 
(blue) colour-composite RGB picture ({\it left}), and WHT continuum-subtracted 
H$_2$ $\lambda$2.122 picture ({\it right}) of NGC\,6772.}
\label{NGC6772}
\end{center}
\end{minipage}
\end{figure*}

The distribution of molecular hydrogen in NGC\,6563 is revealed 
for the first time in our H$_2$ $\lambda$2.122~$\mu$m image (green 
in Figure~\ref{NGC6563}-{\it left} and Figure~\ref{NGC6563}-{\it right}).  
The molecular emission outlines that of the elliptical ionized region, both 
showing a pattern of spiral-like dark lanes and bright filaments.  
These features are typical of bright ring-like PNe such as the Ring 
and the Helix Nebula and have been suggested to form part of a tilted 
barrel-like structure 
\citep{2005AJ....130.1784M,2004AJ....128.2339O,2002AJ....123..346S,2003PASP..115..170S}
The molecular and ionized emissions in the {\it ansae}, however, differ
notably: in [N~{\sc ii}], the emission is uniform, fills the {\it ansae},
and falls off with radial distance, whereas the H$_2$ emission encloses
the [N~{\sc ii}] emission, delineating the {\it ansae} edges with a
remarkable point-symmetric brightness distribution.  
No H$_2$ emission is detected outside the bright inner shell
along the north-south direction, where the walls of this
shell are thicker and may imply more efficient shielding
from the stellar UV radiation.  
On the other hand, weak, diffuse emission is detected forming 
a broken, round shell of radius $\simeq$50\arcsec.  
The emission is notably limb-brightened along an arc towards the east and 
northeast, but it appears fuzzy towards the opposite side of the main nebula.

The emission from the bright optical and H$_2$ shell and {\it ansae} of NGC\,6563 are detected in the IRAC 8~$\mu$m image (Figure~\ref{NGC6563}-{\it left}), revealing more clearly the outer shell of size 105\arcsec\ 
that surrounds the inner elliptical shell and its {\it ansae}.  
This outer shell is well defined along the northeast half,
but its appearance is more diffuse in its southwest half.  
Otherwise, the morphology in the IRAC 8~$\mu$m image of the inner
shell is similar to that of the ionized and hydrogen molecular
material, but there are some subtle differences: the IRAC 8~$\mu$m
image highlights a pattern of filaments inside the elliptical shell
and the {\it ansae} are both broader and extend further.  
Whereas the origin of the IRAC 8~$\mu$m emission of the inner
shell can be attributed to ionized and (most likely) H$_2$
molecular lines, the nature of the material in the outer shell
appears to be H$_2$ emission, although some contribution of thermal 
dust emission cannot be excluded \citep[e.g.,][]{2009MNRAS.399.1126P}.  

\subsubsection{NGC\,6772 --- PN\,G033.1$-$06.3 }\label{NGC6772s}

\begin{figure*}
\begin{minipage}{175mm}   
\begin{center}
\includegraphics[width=80mm]{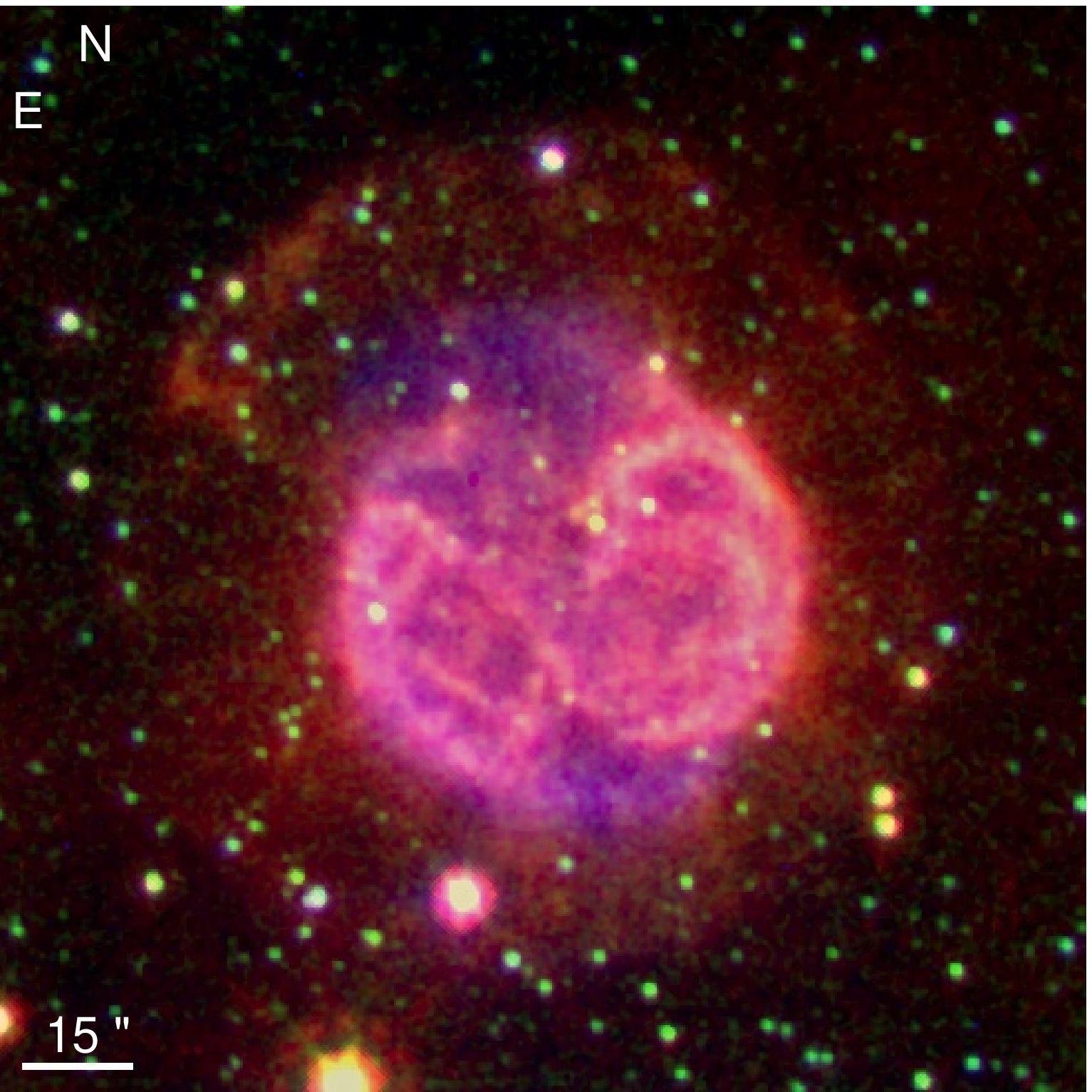}
\hspace*{\columnsep}
\includegraphics[width=80mm]{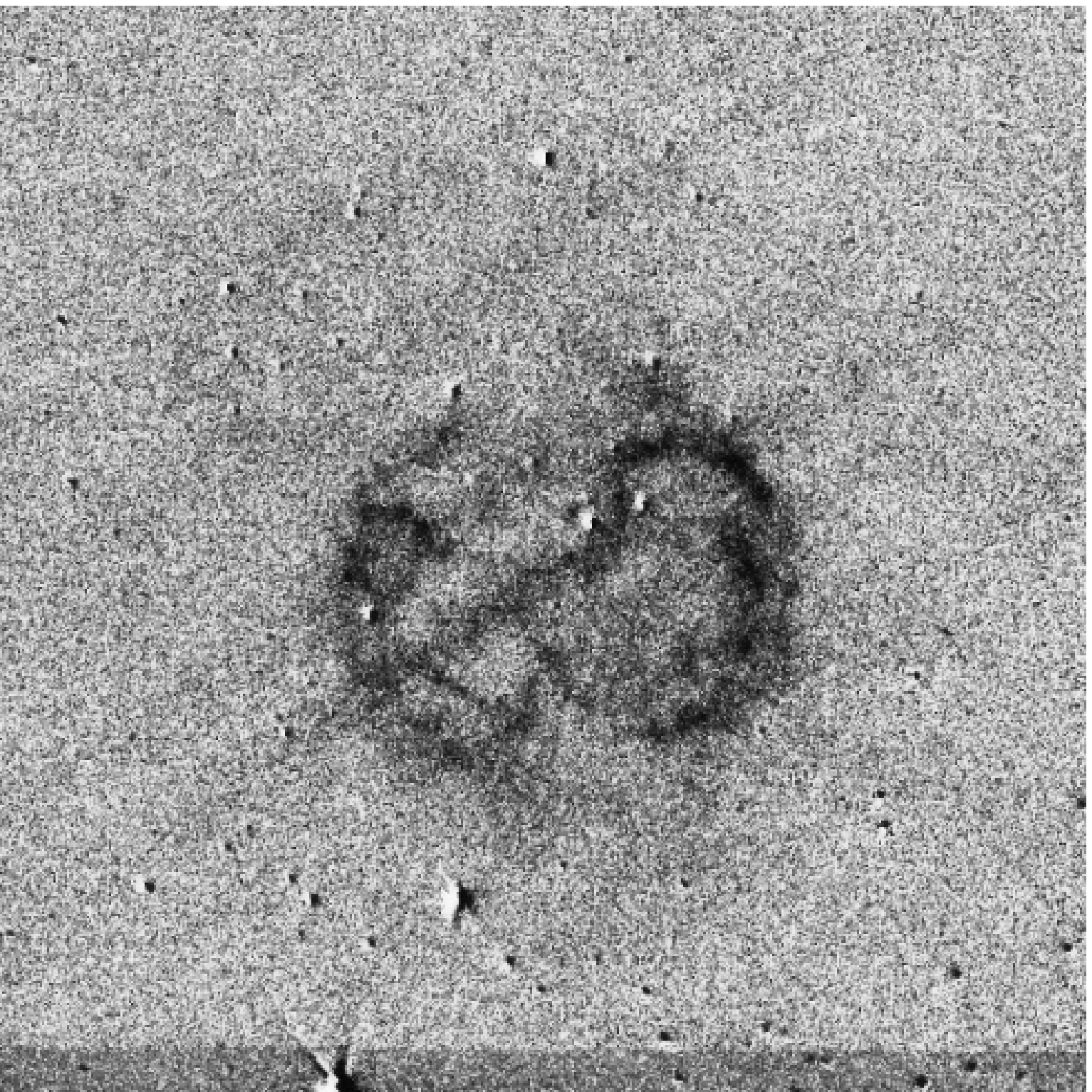}
\caption{
IRAC 8~$\mu$m (red), TNG H$_2$ $\lambda$2.122 (green), and OAN [N~{\sc ii}] 
(blue) colour-composite RGB picture ({\it left}), and TNG continuum-subtracted 
H$_2$ $\lambda$2.122 image ({\it right}) of NGC\,7048. Note that the background-subtracted H$_2$ image is affected of fringing that creates a pattern along the southeast-northwest direction.}
\label{NGC7048}
\end{center}
\end{minipage}
\end{figure*}

NGC\,6772 appears in a large number of optical imaging studies of PNe
\citep{1986ApJ...302..727J,1989AJ.....98.1662J,1992A&AS...96...23S,1993A&A...267..177B,1998ApJS..117..341Z}.
Our [N~{\sc ii}] image (blue in Figure~\ref{NGC6772}-{\it left}) 
confirms the barrel-like elliptical morphology previously described 
\citep[e.g.,][]{1986ApJ...302..727J}. 
The outer edge of this thick elliptical shell, oriented mostly along
the north-south direction, has a size of 31\arcsec$\times$44\arcsec.
The shell is distorted along the northeast-southwest direction,
where two {\it ansae} or blisters protrude from the shell.
Our [N~{\sc ii}] image confirms the presence of outer emission,
mostly distributed along the east-west direction, but it also 
reveals a new structure, an arc of radius $\simeq$66\arcsec\ towards 
the east that appears to be one half of an outer round shell.

Previous studies of the spatial distribution of H$_2$ in NGC\,6772
\citep{1988MNRAS.235..533W} showed an elliptical shell distorted
towards the northeast and southwest regions.
Our H$_2$ $\lambda$2.122 image (Figure~\ref{NGC6772}-{\it right}) reveals a
wealth of details in this elliptical shell, as well as series of features
outside it. The radial features protruding from the inner shell are certainly remarkable.
The arc-like feature hinted in the [N~{\sc ii}] images towards the East of
the main nebular shell is clearly detected in H$_2$ with a similar radius,
$\simeq$66\arcsec.
This structure has a notable limb-brightness morphology towards the east,
whereas it fades and extends further out towards the west.
Overall, this morphology is reminiscent of a shell interacting with
the Interstellar Medium (ISM), either by the nebular proper motion or
by density gradients in the ISM \citep{2009MNRAS.400..575R}.  
Similar morphology can be claimed for NGC\,6563, but the case of 
NGC\,6772 is certainly more clear.

The IRAC 8~$\mu$m emission (red in Figure~\ref{NGC6772}-{\it left}) shows 
the same spatial distribution than the H$_2$ emission in the inner elliptical 
shell, including the distorted regions towards the northeast and southwest.  
The radial features described in H$_2$ are also detected in the 
8~$\mu$m image, but while the H$_2$ rays are concentrated just 
outside the inner elliptical shell, in the IRAC 8~$\mu$m image 
the rays are more evenly distributed around the ellipse and extend 
at greater distances. 
This coincidence can be interpreted as a common origin for this 
emission i.e., molecular hydrogen, whereas the outer section of 
the rays not detected in the H$_2$ image may imply that H$_2$ is 
present but shielded from UV radiation, so that it is not excited 
to emit significantly in the 1--0 S(1) $\lambda$2.122 line.  
The outer shell is more clearly revealed in this IRAC 8 $\mu$m image 
than in the H$_2$ band, with a notable bow-shock morphology towards 
the northeast.  
Along the opposite direction, the IRAC 8~$\mu$m emission is diffuse 
and suggests it is trailing the main nebula that would be moving 
with respect to the ISM.

\subsubsection{NGC\,7048 --- PN\,G088.7$-$01.6}\label{NGC7048s}

Despite being is an extended, relatively bright PN, the morphology and physical 
structure of NGC\,7048 have not been studied in detail.  
The most recent narrow-band optical images mapping its ionized component
were presented by \citet{1987AJ.....94..671B} who described it as a middle 
elliptical PN, a conclusion also reached in other studies based on the 
same sets of images \citep{1998ApJS..117..341Z}.  
Our [N~{\sc ii}] image (blue in Figure~\ref{NGC7048}-{\it left}) shows 
a filamentary, almost round shell of radius $\simeq$30\arcsec\ marked 
by bright eastern and western arcs that leave an opening towards the 
north and south.  
Extended emission is detected along these directions at longer
radial distances, up to $\simeq$40\arcsec\ towards the north and
$\simeq$35\arcsec\ towards the south.  
Rather than an elliptical shell, NGC\,7048 resembles a tilted barrel
that opens at the poles \citep{1993ApJ...404L..25F}.  
Our [N~{\sc ii}] image reveals a weak limb-brightened round  
shell $\simeq$55\arcsec\ in radius that can be described as a 
halo.  
This halo is not completely concentric with the bright inner shell, neither its surface brightness is azimuthally constant: the halo has two bright arcs, north and south of the inner shell, along the directions of their polar openings.

The emission in the H$_2$ $\lambda$2.122~$\mu$m line has been previously 
described by \citet{1996ApJ...462..777K} and \citet{2003MNRAS.344..262D}.  
Both works report bright H$_2$ emission from a filamentary barrel-like
structure, but the larger FoV images of \citet{1996ApJ...462..777K} unveil
emission along P.A.~10\degr, i.e., mostly along the
north-south direction.  
Indeed, this emission is similar to that shown in our [N~{\sc ii}] image.  
A close comparison with our H$_2$ image (Figure~\ref{NGC7048}-{\it right})
confirms these similarities, but it also shows evidences that the diffuse
emission emanating through the north and south openings of the inner shell
is relatively brighter in the [N~{\sc ii}] line (blue color in
Figure~\ref{NGC7048}-{\it left}).  
As the spectroscopic study of \citet{2003MNRAS.344..262D} for the excitation 
mechanism of the H$_2$ emission suggests, the emission from the 
main shell is shock-excited, probably from the propagation of a low
velocity shock generated by the inner shell expanding into the outer
halo \citep{2007apn4.confE..33M}.

The IRAC 8~$\mu$m image (red color in Figure~\ref{NGC7048}-{\it left})
follows the H$_2$ and [N~{\sc ii}] filaments and extended emission of
the inner shell of NGC\,7048, but where these emissions are faintly
detected in the outermost regions, the emission in the IRAC 8~$\mu$m
band is intense and clearly reveals a round shell of radius
$\simeq$56\arcsec.
The emission in this band shows the limb-brightened morphology, 
but it also discloses radial bright and dark stripes and an azimuthally 
dependent brightening.  
These can be associated to the shadowing of the central star by 
the eastern and western arcs of the inner shell that are most 
likely optically thick to the UV radiation from the central star, 
probably in the cooling track of white dwarfs.  
The correspondence between the H$_2$ and [N~{\sc ii}] features and
those in the IRAC 8~$\mu$m image suggests that the emission in the
inner shell and halo detected in this latter band includes significant
contributions of emission lines both from ionized material and molecular 
hydrogen.

\begin{figure*}
\begin{minipage}{175mm}   
\begin{center}
\includegraphics[width=80mm]{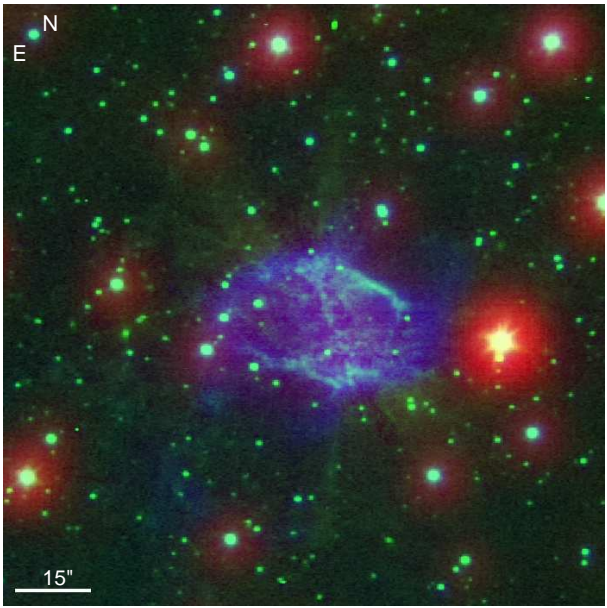}
\hspace*{\columnsep}
\includegraphics[width=80mm]{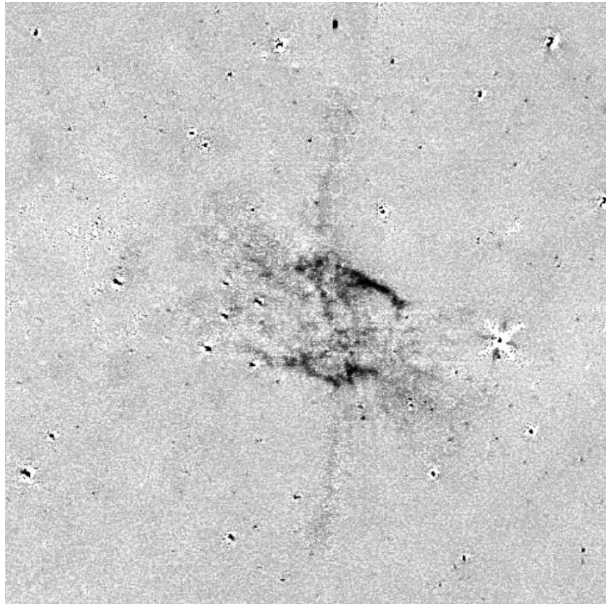}
\caption{
\emph{WISE} W2 4.6~$\mu$m (red), WHT H$_2$ $\lambda$2.122 (green), and OAN 
[N~{\sc ii}] (blue) colour-composite RGB picture ({\it left}), and WHT 
continuum-subtracted H$_2$ $\lambda$2.122 image ({\it right}) of 
M\,1-79.}
\label{M179}
\end{center}
\end{minipage}
\end{figure*}

\subsection{Bipolar PNe}\label{bipolar}

\subsubsection{M\,1-79 --- PN\,G093.3$-$02.4}\label{M179s}

Narrow-band images of M\,1-79 were presented by \citet{1996iacm.book.....M}
and its morphology, kinematics, and physical structure in optical emission 
lines of the ionized gas has been extensively studied by
\citet{1997A&A...326.1187S}.  
This latter work describes M\,1-79 as a 46\arcsec$\times$24\arcsec\ bright
elliptical shell with its major axis oriented near the east-west direction
(P.A.$\simeq$85\degr).  
A bright bar crosses the shell at P.A.$\simeq$14\degr, i.e.,
this bar is misaligned with respect to the ellipse minor
axis and reminds the so called `big-tail' at the central region of the bipolar 
PN NGC\,2818 \citep{2012ApJ...751..116V}.  
A pair of claw-like features protrude from the bright central
region along the southeast-northwest (P.A.$\simeq$140\degr) direction.  
A high contrast image reveals bipolar lobes that extend
further out, up to distances $\simeq$45\arcsec.

Our [N~{\sc ii}] image (Figure~\ref{M179}-{\it left}) confirms these
structural components, additionally revealing that the outer pair of 
bipolar lobes are tilted with respect to the claw-like features and that there 
is an even larger northwestern bipolar lobe which extends up to 
$\simeq$65\arcsec.  
We also detect faint diffuse emission towards the east of the
bright inner shell, but there is no clear evidence it takes
part of a complete outer shell.

The narrow-band H$_2$ $\lambda$2.122 image of M\,1-79 (green in 
Figure~\ref{M179}-{\it left} and Figure~\ref{M179}-{\it right}) discloses 
for the first time the molecular hydrogen distribution in this PN 
which is in many aspects different from the distribution of ionized 
material.  
The outskirts of the optically bright inner shell are delineated
in H$_2$, but there are no signs of the bipolar lobes.  
On the contrary, we detect in H$_2$ a series of bright radial filaments and 
shadows emanating from the bright inner shell that, avoiding the directions 
along which the bipolar lobes are detected, are enclosed within an ellipse 
of size $\simeq$42\arcsec$\times$55\arcsec.   
We note that some of the brightest filaments in H$_2$ are spatially
coincident with the diffuse [N~{\sc ii}] emission detected towards
the east of the bright inner shell.  
We also note that the bright bar in the inner shell produces a
remarkable conical shadow in the H$_2$ emission, thus suggesting
that there is not enough UV flux to excite the
H$_2$ molecules along these directions.

Unfortunately, there are no available \emph{Spitzer} IRAC images of
M\,1-79.  
We have thus used the \emph{WISE} W2 4.6~$\mu$m image to investigate
the properties of this nebula at longer wavelengths.  
This image (red in Figure~\ref{M179}-{\it left}) shows emission from the
central regions of M\,1-79, but the limited spatial resolution and 
sensitivity of \emph{WISE}, and the possible contribution of near-IR ionic 
lines to the W2 band are not adequate to study the molecular component of 
the outer regions of this nebula in this mid-IR image.

\subsubsection{M\,2-48 --- PN\,G062.4$-$00.2 }\label{M248s}

\begin{figure*}
\begin{minipage}{175mm}   
\begin{center}
\includegraphics[width=80mm]{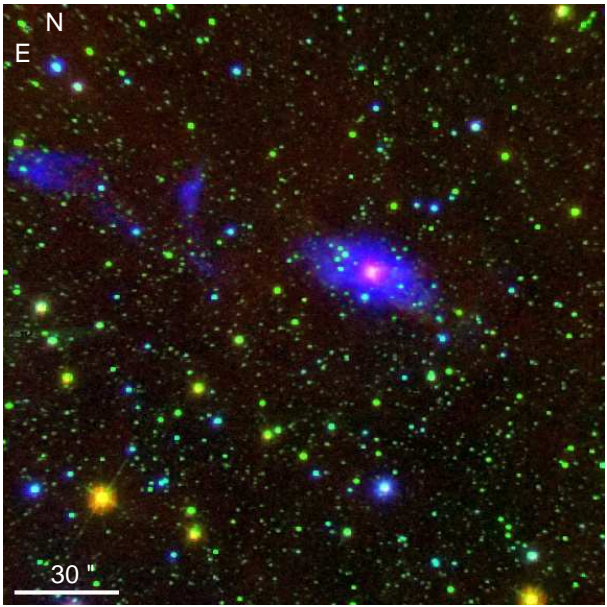}
\hspace*{\columnsep}
\includegraphics[width=80mm]{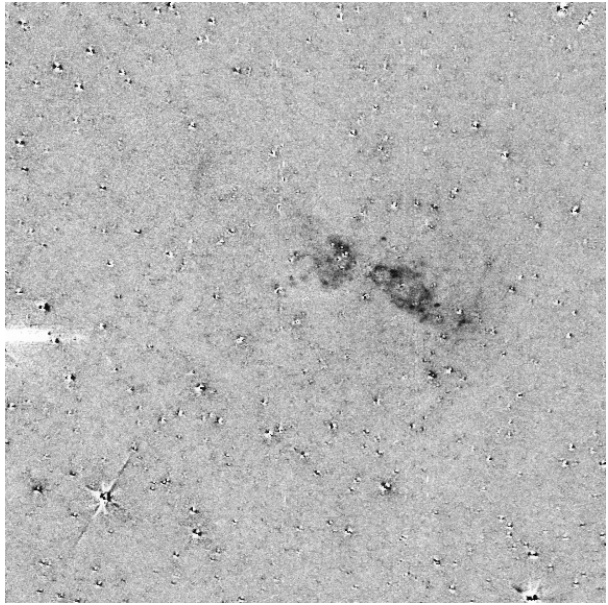}
\caption{
IRAC 8~$\mu$m (red), WHT H$_2$ $\lambda$2.122 (green), and OAN [N~{\sc ii}] 
(blue) colour-composite picture ({\it left}), and WHT continuum-subtracted 
H$_2$ $\lambda$2.122 image ({\it right}) of M\,2-48. 
}
\label{M248}
\end{center}
\end{minipage}
\end{figure*}

Different morphological \citep{1995A&A...293..871C,1996iacm.book.....M} 
and kinematical 
\citep{2000A&A...357.1031V,2002A&A...388..652L,2008AJ....135.2199D} 
studies of M\,2-48 (a.k.a.\ Hen\,2-449) have revealed a highly collimated
bipolar PN with an obscured waist, bow-shock features along its major axis,
and a fragmented off-center round shell.  
Our [N~{\sc ii}] image (blue in Figure~\ref{M248}-{\it left}) confirms
the bow tie-shaped core with size $\simeq$9\arcsec$\times$15\arcsec\ and
detects a bow-shock feature east of the main nebula along the major axis
at $\simeq$55\arcsec, and two outer bow-shock features $\simeq$95\arcsec\
east and $\simeq$120\arcsec\ west along an axis tilted by +5\degr\ with 
respect to the bipolar axis of the main nebula.  
The eastern bow-shock feature at 55\arcsec\ is coincident with
the fragmented round shell that shows a radius of 45\arcsec\ and
is off-centered by 14\arcsec\ towards the northeast. 

Our H$_2$ $\lambda$2.122 image (Figure~\ref{M248}-{\it right}) shows faint
emission encompassing the ionized bipolar lobes, but no H$_2$ emission is
detected at the nebular core.  
There is also even fainter H$_2$ emission just interior of the brightest
eastern and western arcs of the fragmented off-center round shell, and
the eastern bow-shock feature at 55\arcsec.  
The latter is suggestive of shock excitation of the H$_2$ molecules 
at these locations.

The IRAC 8~$\mu$m image of M\,2-48 has been described by
\citet{2008ApJS..174..426K} and \citet{2008MNRAS.383.1029P}.
Here we note that this mid-IR image shows bright
emission at the core of the main nebular shell\footnote{
The $K_c$ image used to obtain the continuum free H$_2$
image in Figure~\ref{M248}-{\it right} also shows bright
emission at the nebular core.},
and faint emission associated to the bipolar lobes.  
The emission from the bipolar lobes in this image has a biconical
morphology and follows more closely the H$_2$ emission than the
[N~{\sc ii}] emission, thus indicating that it corresponds to 
emission from H$_2$ molecules.  
No emission at 8~$\mu$m seems to be associated to the outer round shell
nor to the bow-shock features, but we note that M\,2-48 is embedded within
a region of patchy, diffuse emission.

\subsubsection{NGC\,650-51 --- PN\,G130.9$-$10.5 }\label{NGC650s}

\begin{figure*}
\begin{minipage}{175mm}   
\begin{center}
\includegraphics[width=80mm]{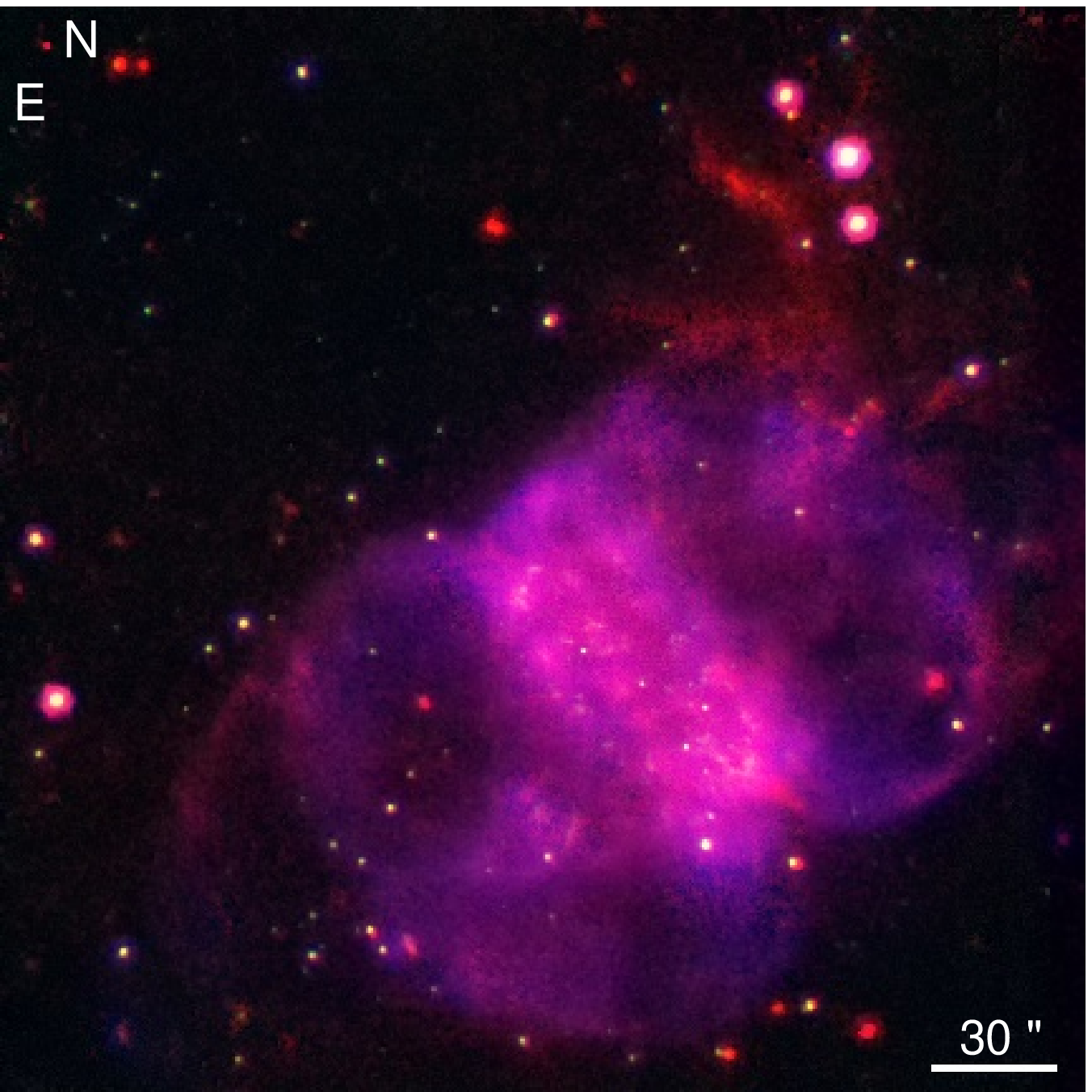}
\hspace*{\columnsep}
\includegraphics[width=80mm]{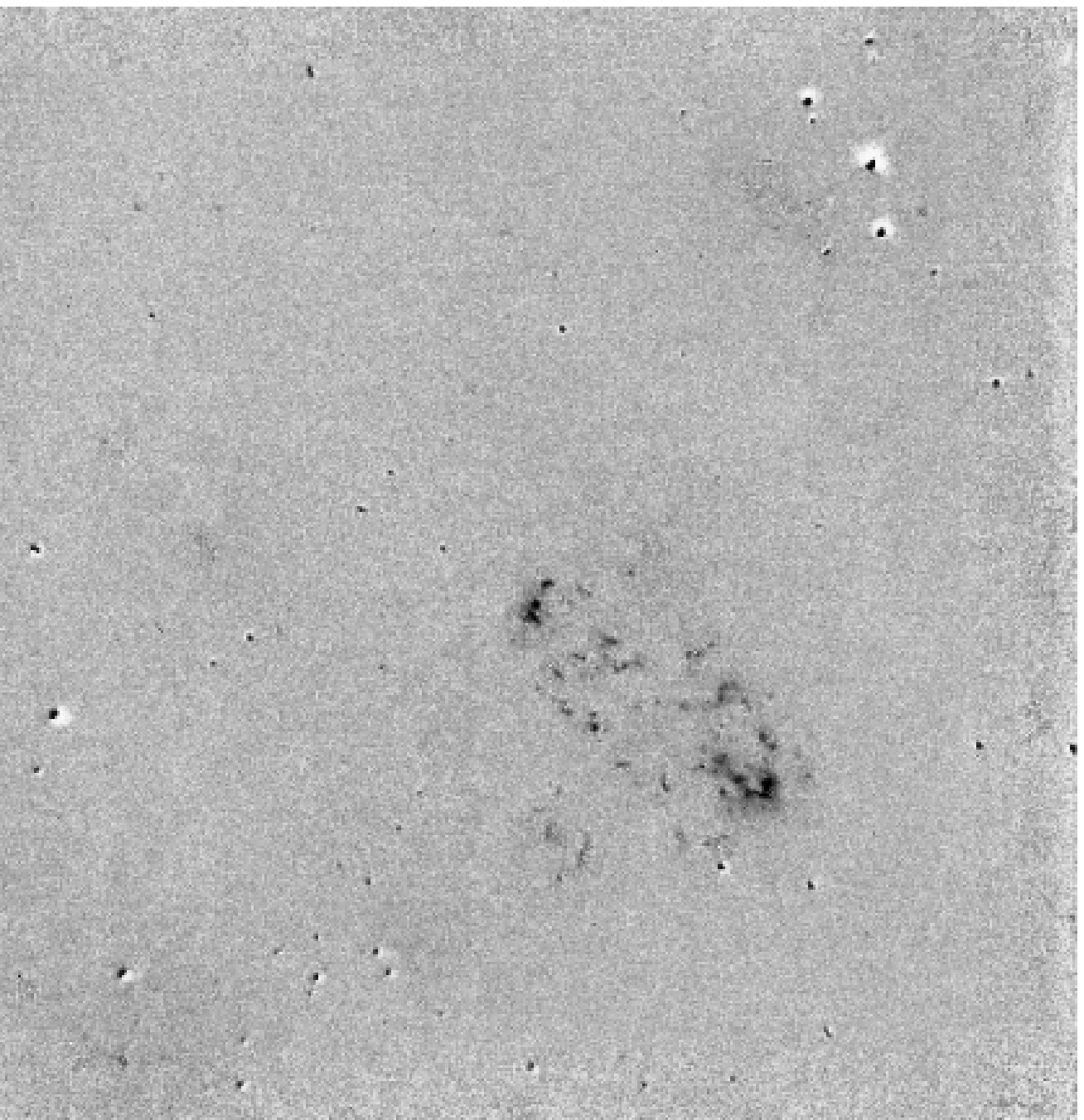}
\caption{
IRAC 8~$\mu$m (red), WHT H$_2$ $\lambda$2.122 (green), and OAN 
[N~{\sc ii}]+H$\alpha$ (blue) colour-composite RGB picture ({\it left}), 
and WHT continuum-subtracted H$_2$ $\lambda$2.122 image ({\it right}) of 
NGC\,650-51.}
\label{NGC650}
\end{center}
\end{minipage}
\end{figure*}

The optical and mid-IR properties of NGC\,650-51 (a.k.a.\ M\,76) 
have been recently studied in detail by \citet[][see also references 
therein]{2008MNRAS.391...52R}.  
Our [N~{\sc ii}] image (blue in Figure~\ref{NGC650}-{\it left}) shows a 
bright, tilted ring-like structure oriented along the northeast to southwest direction 
with an angular size of 66\arcsec$\times$144\arcsec.  
Two bipolar lobes with archetypical butterfly morphology extend along 
the minor axis of the central ring up to distances $\simeq$85\arcsec.  
At the tips of these lobes, fainter bow-shaped structures are 
detected, increasing the total extent of the bipolar lobes up to 260\arcsec.

Previous observations of NGC\,650-51 in the H$_2$ line emphasized the 
presence of diffuse emission from the central ring, which seemed to 
be brighter at the tips of its long axis \citep{1996ApJ...462..777K}.  
Our H$_2$ image (Figure~\ref{NGC650}-{\it right}) resolves this ring into a 
series of disconnected knots and filaments.  
Some knots and filaments are also detected in the inner 
regions of the bipolar lobes.

The spatial distribution of the emission of NGC\,650-51 in the IRAC 8 $\mu$m 
image (red in Figure~\ref{NGC650}-{\it left}) has been compared to those of 
the ionized and H$_2$ material 
\citep[see also][]{2004ApJS..154..296H,2008MNRAS.391...52R}.  
The emission in this mid-IR band closely follows the [N~{\sc ii}] image, 
showing the central ring, the inner bipolar lobes, and their fainter 
extensions.  
In addition, the mid-IR emission highlights faint filaments outside 
the main nebular body.   

\subsubsection{NGC\,6537 --- PN\,G010.1$+$00.7 }\label{NGC6537s}

NGC\,6537 is a high velocity bipolar PN with a noticeable point-symmetric
brightness distribution of the bipolar lobes \citep{1993A&A...269..462C}.
Narrow-band images obtained with the \emph{Hubble Space Telescope}
confirm these morphological features and also reveal a dust shell
at its center that is suspected to collimate the bipolar lobes
\citep{2005MNRAS.363..628M}.
This morphological description is supported by our [N~{\sc ii}] image 
(blue in Figure~9-{\it left}), in which two bipolar lobes, extending up to 
2\arcmin\ from the center along the northeast-southwest direction, can be noticed.

\begin{figure*}
\begin{minipage}{175mm}   
\begin{center}
\includegraphics[width=80mm]{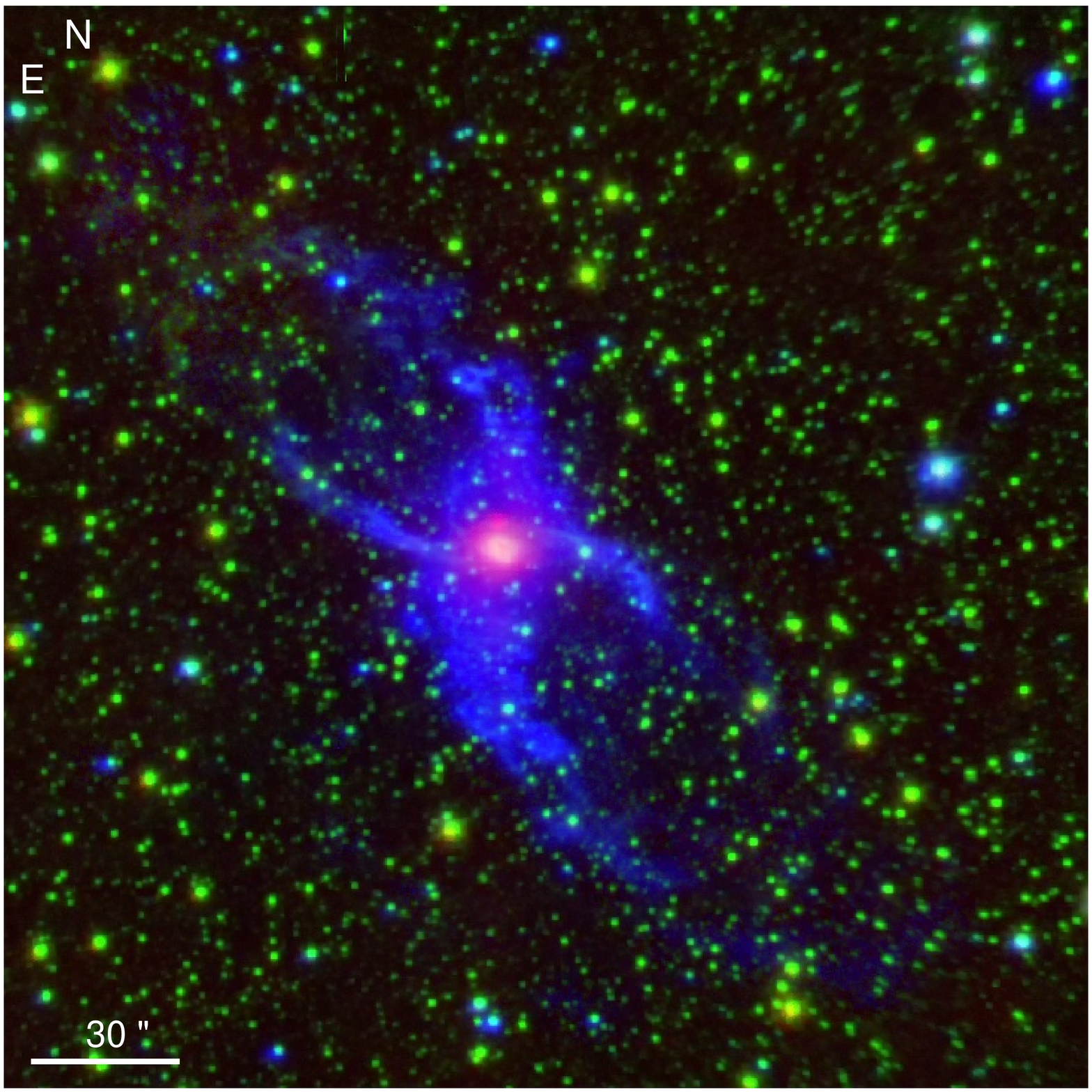}
\hspace*{\columnsep}
\includegraphics[width=80mm]{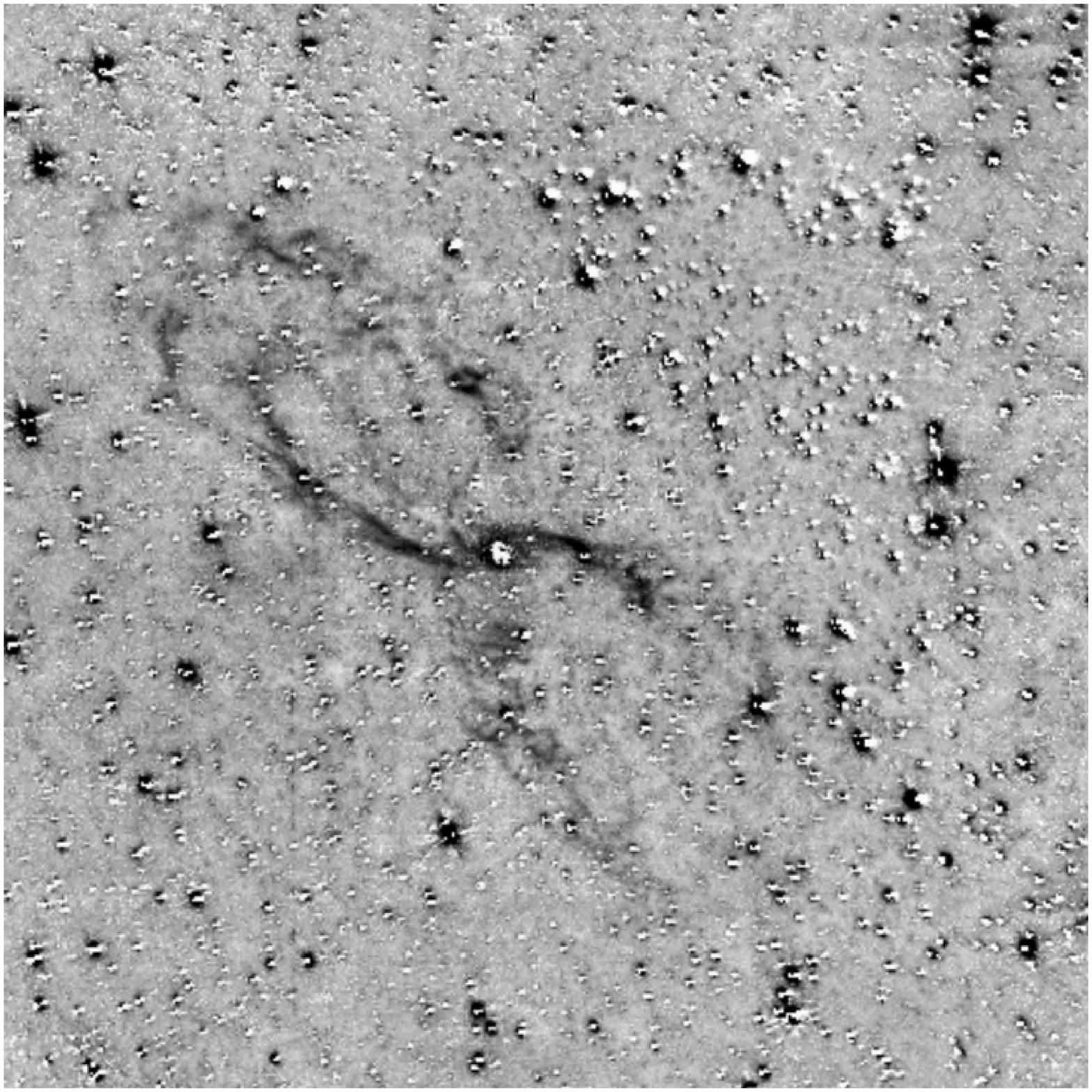}
\caption{
IRAC 8~$\mu$m (red), WHT H$_2$ $\lambda$2.122 (green), and OAN [N~{\sc ii}] 
(blue) colour-composite RGB picture (left), and WHT continuum-subtracted H$_2$ 
$\lambda$2.122 image ({\it right}) of NGC\,6537.}
\label{NGC6537}
\end{center}
\end{minipage}
\end{figure*}

The narrow-band H$_2$ images of NGC\,6537 presented by
\citet{1996ApJ...462..777K} and \citet{2003MNRAS.344..262D}
show strong emission at the nebular core and along an S-shaped
line that follows the point-symmetric distribution of the
limb-brightened edges of the bipolar lobes.
The detection of H$_2$ emission at the nebular core is uncertain,
as $K_c$ images also show bright emission at this location, but
the images presented by \citet{2003MNRAS.344..262D} seem to
confirm a ring-like feature of H$_2$ emission.
Our continuum-subtracted H$_2$ image (Figure~\ref{NGC6537}-{\it right})
displays this ring, but it also traces the faintest emission from the
bipolar lobes.
Relatively strong H$_2$ emission is detected at the tip of the
northeastern bipolar lobe which is otherwise rather faint in
the [N~{\sc ii}] image.

The IRAC 8~$\mu$m image of NGC\,6537
\citep[red in Figure~\ref{NGC6537}-{\it left},
see also][]{2008ApJS..174..426K,2008MNRAS.383.1029P} shows a bright
unresolved source at the nebular core and faint diffuse emission
that traces the inner regions of the bipolar lobes.
At least for the northeastern lobe, its tip is detected in the 8~$\mu$m
band.

\subsubsection{NGC\,6778 --- PN\,G034.5$-$06.7 }\label{NGC6778s}

\begin{figure*}
\begin{minipage}{175mm}   
\begin{center}
\includegraphics[width=80mm]{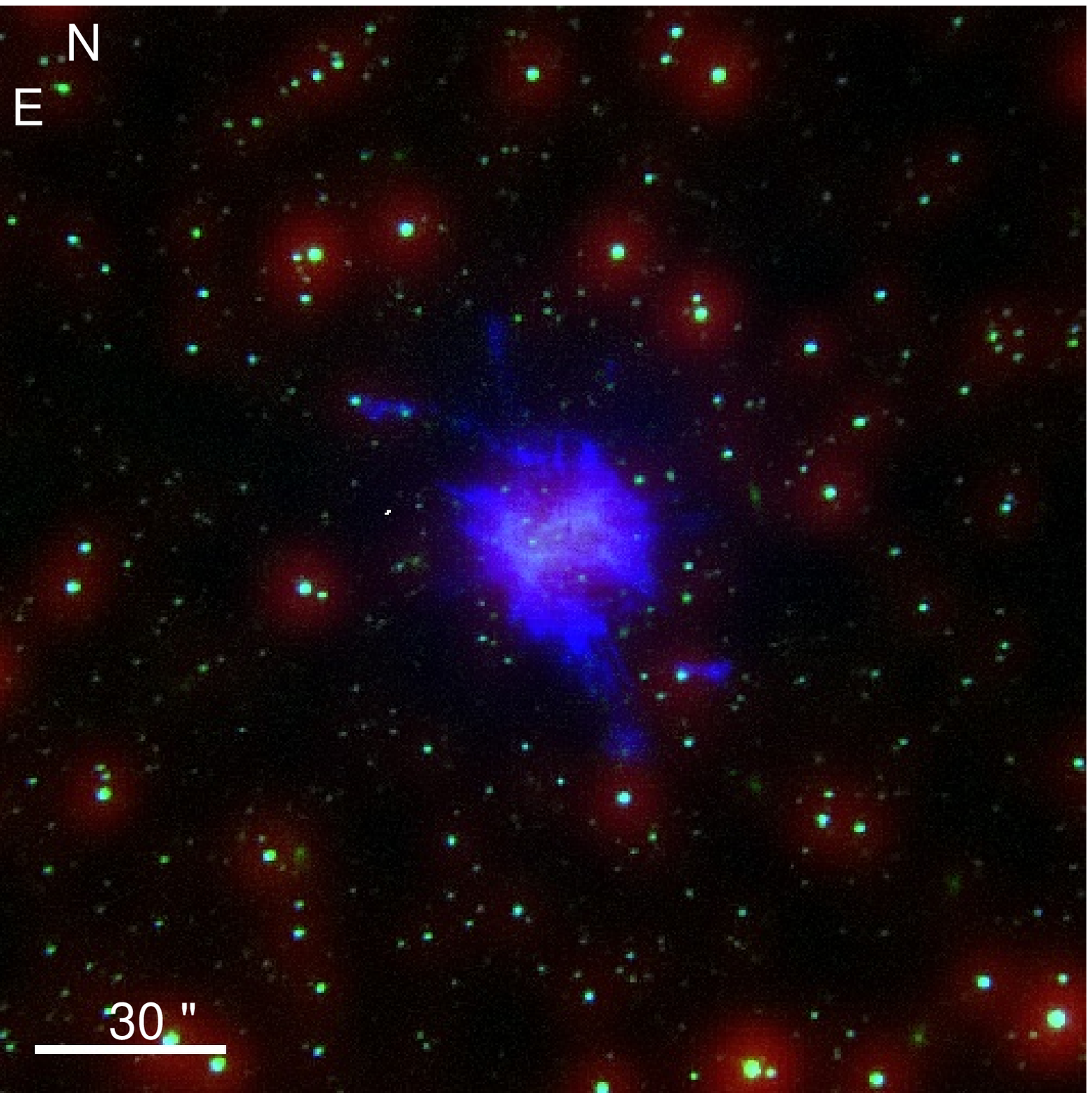}
\hspace*{\columnsep}
\includegraphics[width=80mm]{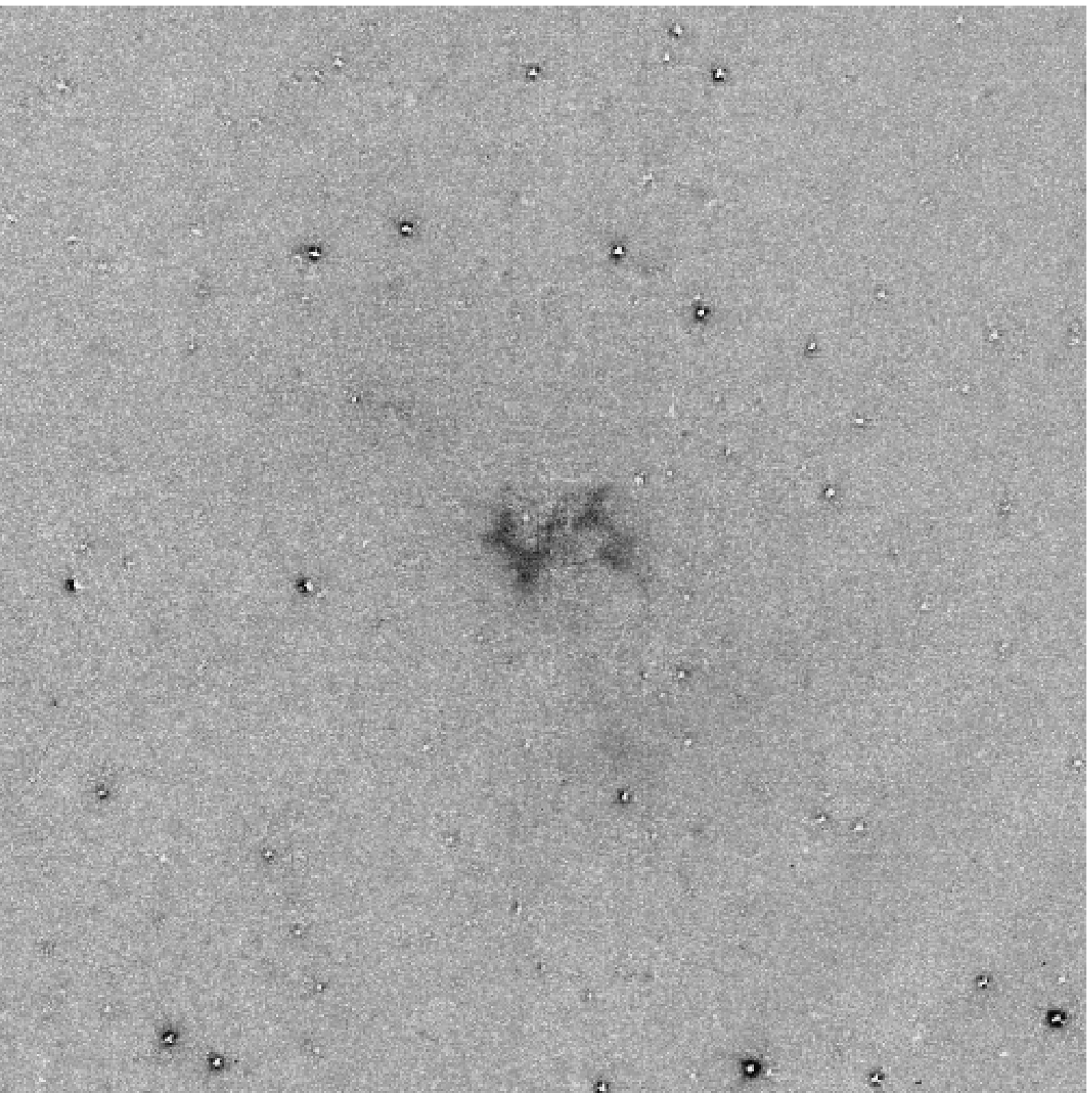}
\caption{
\emph{WISE} W2 4.6~$\mu$m (red), WHT H$_2$ $\lambda$2.122 (green), and 
NOT [N~{\sc ii}] (blue) colour-composite RGB picture ({\it left}), and 
WHT continuum-subtracted H$_2$ $\lambda$2.122 image ({\it right}) of 
NGC\,6778. 
}
\label{NGC6778}
\end{center}
\end{minipage}
\end{figure*}

NGC\,6778 had received little attention until the discovery of a
binary CSPN \citep{2011A&A...531A.158M} and a disrupted equatorial
ring fragmented by fast stellar winds and multiple collimated
outflows \citep{2012A&A...539A..47G}.
The [N~{\sc ii}] image \citep[blue in Figure~\ref{NGC6778}-{\it left}, adopted
from][]{2012A&A...539A..47G} has been compared to the H$_2$ $\lambda$2.122
$\mu$m (green in Figure~\ref{NGC6778}-{\it left}) and continuum-subtracted
H$_2$ (Figure~\ref{NGC6778}-{\it right}) images.
The H$_2$ emission traces the brightest [N~{\sc ii}] emission at the 
tips of the equatorial regions.  
This spatial distribution is reminiscent of a barrel-like structure and 
may imply that the object is density-bounded along the equatorial 
plane.

Unfortunately, there are no available \emph{Spitzer} images of NGC\,6778.
The \emph{WISE} W2 4.6 $\mu$m image (red in Figure~\ref{NGC6778}-{\it left})
shows a bright, unresolved source at the location of the central regions
of NGC\,6778.
The limited spatial resolution and sensitivity of \emph{WISE}, and the
possible contribution of near-IR ionic lines to the W2 band are not
adequate to study the molecular component of the outer regions of this
nebula.

\section{Discussion}\label{sec_dis}

\subsection{Interpreting the emission in IRAC 8 $\mu$m images}

Multiple studies have shown the efficiency of IRAC images in the 8~$\mu$m 
band (6.2994~$\mu$m $\leq \lambda \leq$9.5876~$\mu$m) to detect extended
haloes around PNe \citep[e.g.,][]{2009MNRAS.400..575R} and dense knots
embedded within ionized nebular shells \citep[e.g.,][]{2006ApJ...652..426H}.
This emission can be attributed to the contribution into the IRAC 8~$\mu$m
bandpass of the H$_2$ 0--0 S(5) $\lambda$6.9091 $\mu$m, 1-1 S(5)
$\lambda$7.297~$\mu$m, and 0--0 S(4) $\lambda$8.0258 $\mu$m rotational
lines.
However, the contribution of other emission lines from ionic species, such as
[Ar~{\sc ii}] $\lambda$6.985 $\mu$m,
[Ne~{\sc iv}] $\lambda$7.642 $\mu$m,
[Ar~{\sc v}] $\lambda$7.902 $\mu$m, and
[Ar~{\sc iii}] $\lambda$8.991 $\mu$m, cannot be neglected for
the inner ionized nebular regions \citep{2004ApJS..154..296H}.
Similarly, the contribution of the PAH features at 6.2, 7.7 and 8.6~$
\mu$m may be of importance for dusty regions, such as obscured
equatorial waists of bipolar PNe and dense knots.

The comparison between H$_2$ and IRAC 8 $\mu$m images of PNe in our sample 
reveals a close correlation between morphological features seen in the two 
bands for a significant fraction of sources.
The central regions and outermost shells of A\,66, NGC\,6563,
NGC\,6772, and NGC\,7048 show overall shapes and detailed 
morphological features which are very similar in both bands.
The H$_2$ and IRAC 8~$\mu$m images of A\,66 display a fragmentary
ring and a series of small-scale structures with an appearance of
cometary knots.
These knots, whose heads are clearly detected in the innermost regions
of A\,66 mapped by the H$_2$ image, extend further out in long filaments
that are very notable in the larger FoV of the IRAC 8~$\mu$m image.
Similarly, the inner nebulae of NGC\,6563, NGC\,6772, and NGC\,7048 show
an excellent correspondence in the H$_2$ and 8~$\mu$m images, even on
small-scale filaments and knots.
The H$_2$ images of these PNe are indicative of limb-brightened, faint
outer shells that are brighter and revealed as complete envelopes in
the IRAC 8~$\mu$m images.
These shells, which can be described as haloes \citep{1987ApJS...64..529C}, are spatially coincident in
the H$_2$ and 8~$\mu$m images.
As for A\,66, there is a series of bright (rays) and dark (shadows) radial
filaments that connect the inner and outer shells of NGC\,6772.

The similar spatial distribution of the H$_2$ and IRAC 8~$\mu$m images in these PNe suggests that a significant fraction (if not all) of the emission in the IRAC 8~$\mu$m images of these PNe can be attributed to lines of molecular hydrogen in this IRAC filter bandpass. This conclusion is supported by the \emph{Spitzer} IRS spectra of NGC\,6720 and NGC\,7293 which show that the IRAC 8~$\mu$m is dominated by H$_2$ emission lines \citep{2005AAS...206.3901H,2006ApJ...652..426H,2009eimw.confE..29H}. To test this suggestion, we have examined the \emph{Spitzer} IRS spectra in the 5.8-8.0 $\mu$m band available for the PNe in our sample, namely M\,2-51 and NGC\,6537. The spectra presented in Figure~11 shows that the mid-IR emission in the IRAC 8~$\mu$m band from M\,2-51 and the bipolar lobes of NGC\,6537 present prominent H$_2$ emission of the transitions 0--0 S(5) $\lambda$6.9091~$\mu$m, 1--1 S(5) $\lambda$7.2801~$\mu$m, and 0--0 $\lambda$8.2058~$\mu$m. On the other hand, the mid-IR emission from the central regions of NGC\,6537 is dominated by emission lines of ionic species such as [Ne~{\sc iv}] $\lambda$7.642~$\mu$m and [Ar~{\sc iii}] $\lambda$8.991 $\mu$m, and the PAH feature at 7.7~$\mu$m. 

We are thus confident that the extended, outermost emission detected in the IRAC 8 $\mu$m images can be attributed to H$_2$, whereas some contribution of emission lines from ionized material can be expected in the innermost regions.  The H$_2$ emission in the PNe in our sample is mostly associated to shell-like structures and its excitation may be two-fold as discussed below.

The H$_2$ emission is mostly associated to shell-like structures and
its excitation may be two-fold as discussed below.
The inner shell of NGC\,7048 is shock-excited \citep{2003MNRAS.344..262D},
probably associated to the propagation of a small-velocity shock
generated by the expansion of the inner shell into the outer shell.
The emission of H$_2$ in the inner shells of NGC\,6563 and NGC\,6772
may be similarly shock-excited, whereas the H$_2$ emission from the 
outer shells of A\,66, NGC\,6772 and NGC\,7048 seems to exhibit a 
dependence with ``openings'' in the inner shell that is suggestive 
of UV excitation.

Contrary to these nebulae, the spatial distributions of the 
H$_2$ and IRAC 8~$\mu$m emissions in the bipolar PNe M\,2-48, NGC\,650-51, 
and NGC\,6537, do not correlate closely.
The H$_2$ emission mainly traces the bipolar lobes of M\,2-48 and NGC\,6537,
but their IRAC 8~$\mu$m images reveal bright emission at their cores, with
faint emission from the bipolar lobes.
In sharp contrast, the H$_2$ emission of NGC\,650-51 traces its equatorial
torus and knotty features in the bipolar lobes, but bright IRAC 8~$\mu$m
emission also arises from the bipolar lobes

The lack of \emph{Spitzer} IRAC observations does not allow us to 
study the relative spatial distributions of H$_2$ and 8~$\mu$m 
emission for M\,1-79 and NGC\,6778.
As for NGC\,6772 and NGC\,7048, the radial features of H$_2$ emission
seen in M\,1-79 are indicative of UV excitation and shielding effects.
This may also be the case for the knotty appearance of the equatorial
ring of NGC\,6778 seen in H$_2$.

\subsection{Origin of the hydrogen molecular material}

The spatial correspondence between H$_2$ and 8~$\mu$m emission can
be interpreted by the contribution of H$_2$ emission lines into the
IRAC 8~$\mu$m bandpass.
Alternatively, the emission in this IRAC band may be attributed to thermal
continuum emission from dust coexisting with molecular hydrogen.
The spatial coincidence of molecular hydrogen and dust can have important
consequences for the origin of the molecular material.
The dust may act as a shield for H$_2$ molecules from the ionizing UV
radiation of the star, and thus dust and H$_2$ may have been present
in the nebulae since its formation.
Alternatively, the dust grains may act as catalyst for the formation
of new H$_2$ molecules on their surface.
These phenomena have been studied by \citet{2009ApJ...700.1067M} for
NGC\,7293 (Helix Nebula) and \citet{2010A&A...518L.137V}
for NGC\,6720 (Ring Nebula).

The survival of dense, dusty knots, formed during the AGB phase within 
the ionized zone is critical for the origin of the molecular material 
\citep{2003MNRAS.345.1291R,2006ApJ...646L..61G}.  
If the knots are able to survive a long time, then it can be expected
that coeval H$_2$ has survived shielded from the UV stellar radiation 
by the high density and relatively low temperature provided by 
the knots.  
However, if these dense knots are rapidly destroyed by UV radiation, then 
the H$_2$ material detected in old PN should have condensated onto newly 
formed dust grains.

\subsection{H$_2$ emission and nebular morphology} 

The literature provides a wealth of observational evidence supporting the
prevalence of H$_2$ emission among bipolar PNe with respect to other
morphological types \citep[][and references therein]{1996ApJ...462..777K}.  
Bipolar PNe seem to possess important reservoirs of molecular material,
either because they descend from more massive progenitors and have therefore
more massive envelopes, or either because the bipolar geometry provides a
suitable haven for the survival of molecules in dense equatorial regions.  
Furthermore, bipolar PNe seem to offer suitable excitation conditions
for the excitation and emission of the H$_2$ molecule, may be through 
shock-excitation, but most likely by offering an appropriate flux of UV 
exciting photons as the H$_2$ emission is associated in most cases to UV 
excitation \citep{2006AJ....131.1515L}.  
The quick evolution of the central star of a bipolar planetary nebula --assuming it descends
from a massive progenitor--, implies it reaches the high effective temperature
necessary to provide a suitable UV flux of photons in a short time-scale
\citep{2004ApJ...607..865A}.  
The combined effects of post-AGB speed evolution and nebular geometry
may indeed play an important role, as bipolar PNe that exhibit an
equatorial ring structure have much stronger H$_2$ emission than bipolar
PNe with a narrow waist \citep{2000ApJS..127..125G}.

The prevalence of H$_2$ emission among bipolar PNe led to postulate the
so-called {\it Gatley's rule} \citep{1996ApJ...462..777K} stating that
``the detection of the 2.122 $\mu$m S(1) line of H$_2$ is sufficient to
determine the bipolar nature of a PN.''  
This conclusion was based on the correlation between H$_2$ detection
and bipolar morphology of a sample of PNe, although some of the PNe
in that sample exhibiting H$_2$ emission are not strictly bipolar,
i.e., they do not show a butterfly morphology or bipolar lobes
connected by an equatorial ring or a waist
\citep{1987AJ.....94..671B,1995A&A...293..871C,2000ASPC..199...17M}.
For example, the physical structure of NGC\,6720, the Ring Nebula, has
been controverted and sometimes assumed to be bipolar, but a recent study
by \citet{2007AJ....134.1679O}
confirms the closed ellipsoidal shape of the inner shell proposed by
\citet{1997ApJ...487..328G}.  

\begin{figure}
\begin{center}
\includegraphics[viewport= 90 0 554 600,clip,angle=0,width=0.6\textwidth]{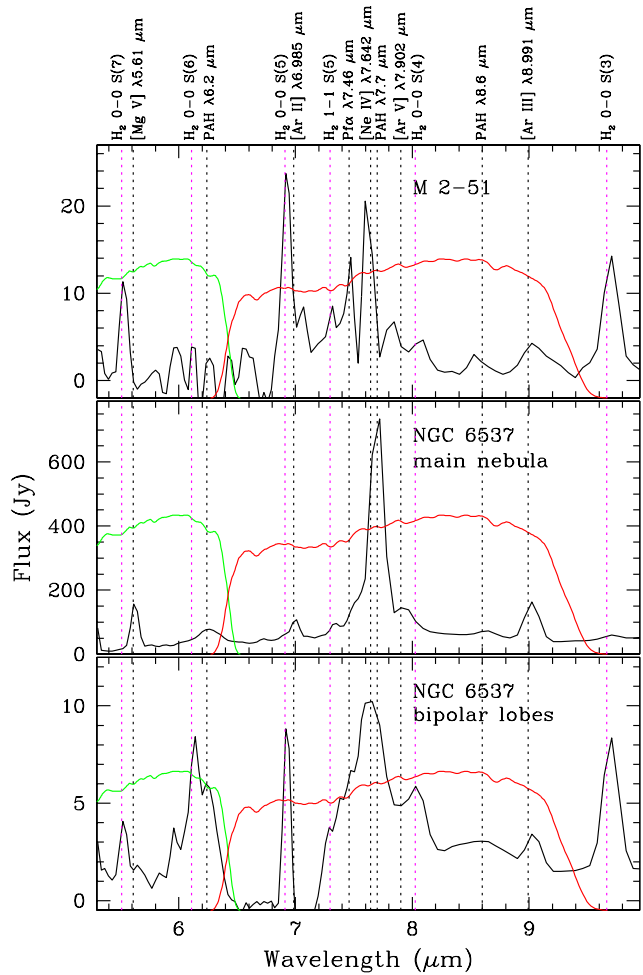}
\end{center}
\caption{
\emph{Spitzer} IRS SL spectra of peripheral region of M\,2-51 ({\it top}) and main nebula region ({\it center}), and bipolar lobes ({\it low}) of NGC\,6537. The spectra show the IRAC 5.8~$\mu$m ({\it green}) and 8~$\mu$m ({\it red}) response profiles. Multiple H$_2$ and ionic lines, and some PAHs bands are marked. There is no definite identification for the bright feature at $\sim$7.6 $\mu$m, especially in the spectrum of the innermost regions of NGC\,6537. We note that this is the wavelength at which the SL1 and SL2 spectra joins, thus we cannot discard it is a spurious artifact. 
}
\label{spec}
\end{figure}

The detection of H$_2$ emission from PNe in our sample with shell-like
morphologies (A\,66, M\,2-51, NGC\,6563, NGC\,6772, and NGC\,7048) 
seems to violate {\it Gatley's rule}.  
We concur that some of these PNe can be described as ellipsoidal shells
with bipolar extensions (e.g., M\,1-79), or barrel-like structures with
shorter extensions or {\it ansae} (e.g., NGC\,6563 and NGC\,7048).  
The kinematical information available for some of them in the literature
or in the SPM catalogue \citep{2012RMxAA..48....3L},
however, implies that M\,2-51, NGC\,6563, NGC\,6772, and NGC\,7048 cannot
be described by no means as bipolar PNe.  
Similarly, the detection of H$_2$ emission from haloes in A\,66,
NGC\,6563, NGC\,6772, and NGC\,7048 \citep[and probably some more in
the literature, e.g.,][]{2009MNRAS.399.1126P,2009MNRAS.400..575R}
does not conform {\it Gatley's rule.}

Even among the bipolar PNe in our sample, we appreciate notable differences.  
Sources that do not have an equatorial ring (M\,2-48 and NGC\,6537) show
bright 8~$\mu$m emission at their cores, but the H$_2$ emission arises
mostly from the bipolar lobes.  
On the other hand, the H$_2$ emission from sources with an equatorial ring
(NGC\,650-51 and NGC\,6778) originates from these equatorial regions.  
In these sources, we note that the H$_2$ emission does not arise from
a torus external to the ionized one, but from dense clumps or knots
embedded within the ionized ring.  
This situation is similar to the H$_2$ emission 
detected in NGC\,6720 and NGC\,7293 
\citep{2006IAUS..234..173H,2006ApJ...652..426H,2009ApJ...700.1067M,2010A&A...518L.137V,2002AJ....123..346S}, 
and reminiscent of the knots that occupy the 
whole volume of the main nebula of NGC\,6853 
\citep{2007apn4.confE...7M} or those that we 
detect in A\,66.

Contrary to previous interpretations, the presence of a thick equatorial 
structure in bipolar PNe does not imply H$_2$ emission: such structures 
may provide a haven for the survival of hydrogen molecules, but at the 
same time UV radiation cannot excite these molecules, and thus H$_2$ 
emission is not produced.  
Meanwhile the H$_2$ emission from tori of bipolar PNe seems to come 
from knots that shield themselves from the UV radiation of the central star.  


\section{Summary}

We have compared the emission detected in IRAC 8~$\mu$m and near-IR H$_2$ images 
to investigate the nature of the emission observed in this mid-IR IRAC band in a 
sample of PNe.  
We confirm that a significant fraction of the IRAC 8~$\mu$m emission can be 
attributed to H$_2$ line emission, thus revealing the molecular nature of 
the material seen in these IRAC images.  
The H$_2$ emission arises from inner shells and outer envelopes or haloes 
of round and elliptical PNe, as well as from bipolar lobes and dense knots in the 
equatorial rings of bipolar PNe. 
We found that H$_2$ emission is not exclusively associated to bipolar 
PNe, but objects with a barrel-like physical structure and their haloes 
have also important amounts of molecular hydrogen. 
We also suggest that the H$_2$ emission from equatorial rings of bipolar 
PNe arises from discrete knots, rather than from a photo-dissociation region 
just exterior to the ionized ring.

\section*{Acknowledgments}

RAML acknowledges support from CONACyT by the CVU~79367 programs ``Becas Nacionales'' and ``Becas Mixtas de Movilidad en el Extranjero''. 
He also acknowledges the Instituto de Astrof\'isica de Andaluc\'ia for its great hospitality and the facilities provided for the realization of this work.
GRL acknowledges support from CONACyT (grant 177864) and PROMEP (Mexico).
RV, MAG, and GRL thank support by grant IN109509 (PAPIIT-DGAPA-UNAM). 
MAG and GRL acknowledges partial support of the Spanish grants AYA~2008-01934 and AYA~2011-29754-C03-02 of the Spanish Ministerio de Ciencia e Innovaci\'on (MICINN) and Ministerio de Econom\'ia y Competitividad (MEC) which includes FEDER funds. 

Paper based in part on ground-based observations from: the Observatorio Astron\'omico Nacional at the Sierra de San Pedro M\'artir (OAN-SPM), which is a national facility operated by the Instituto de Astronom\'{\i}a of the Universidad Nacional Aut\'onoma de M\' exico; the Italian Telescopio Nazionale Galileo (TNG) operated on the island of La Palma by the Fundaci\'on Galileo Galilei of the INAF (Istituto Nazionale di Astrofisica) at the Spanish Observatorio del Roque de los Muchachos of the Instituto de Astrof\'isica de Canarias; the William Herschel Telescope, operated on the island of La Palma by the Isaac Newton Group in the Spanish Observatorio del Roque de los Muchachos of the Instituto de Astrof\'isica de Canarias; the Nordic Optical Telescope, operated on the island of La Palma jointly by Denmark, Finland, Iceland, Norway, and Sweden, in the Spanish Observatorio del Roque de los Muchachos of the Instituto de Astrof\'isica de Canarias; and the New Technology Telescope at the La Silla Observatory.

This publication makes use of data products from the Two Micron All Sky Survey, which is a joint project of the University of Massachusetts and the Infrared Processing and Analysis Center/California Institute of Technology, funded by the National Aeronautics and Space Administration and the National Science Foundation.

Based in part on photographic data obtained using The UK Schmidt Telescope. The UK Schmidt Telescope was operated by the Royal Observatory Edinburgh, with funding from the UK Science and Engineering Research Council, until 1988 June, and thereafter by the Anglo-Australian Observatory. Original plate material is copyright (c) of the Royal Observatory Edinburgh and the Anglo-Australian Observatory. The plates were processed into the present compressed digital form with their permission. The Digitized Sky Survey was produced at the Space Telescope Science Institute under US Government grant NAG W-2166.

This work is based in part on observations made with the Spitzer Space Telescope, which is operated by the Jet Propulsion Laboratory, California Institute of Technology under a contract with NASA.

This publication makes use of data products from the Wide-field Infrared Survey Explorer, which is a joint project of the University of California, Los Angeles, and the Jet Propulsion Laboratory/California Institute of Technology, funded by the National Aeronautics and Space Administration.

\bsp

\label{lastpage}

\end{document}